\documentclass{aa}

\usepackage{natbib}
\usepackage[breaklinks=true]{hyperref} 
\bibpunct{(}{)}{;}{a}{}{,} 

\usepackage{xspace}

\makeatletter
\renewcommand*\env@matrix[1][\arraystretch]{%
  \edef\arraystretch{#1}%
  \hskip -\arraycolsep
  \let\@ifnextchar\new@ifnextcharcan
  \array{*\c@MaxMatrixCols c}}
\makeatother

\usepackage{graphicx}	

\usepackage{color}
\usepackage{enumitem}
\usepackage{pdflscape}
\usepackage[varg]{txfonts}
\usepackage{amsmath}	

\begin{document}

\title{Chromatically modelling the parsec scale dusty structure in the centre of NGC\,1068\thanks{This work makes use of the following ESO programmes: 0103.C-0143, 0104.B-0322, 0102.B-0667, 0102.C-0205 and 0102.C-0211}}

\titlerunning{Chromatic modelling of NGC\,1068}

\author{J. H. Leftley
        \inst{1}
        \and 
        R. Petrov \inst{1}
        \and
        N. Moszczynski\inst{2}
        \and
        P. Vermot\inst{2}
        \and
        S. F. H\"{o}nig\inst{3}
        \and
        V. Gamez Rosas \inst{4}
        \and
        J. W. Isbell\inst{5}
        \and
        W. Jaffe\inst{4}
        \and
        Y. Cl\'enet\inst{2}
        \and
        J.-C. Augereau\inst{6}
        \and
        P. Berio\inst{1}
        \and
        R. I. Davies\inst{7}
        \and
        T. Henning\inst{5}
        \and
        S. Lagarde\inst{1}
        \and
        B. Lopez\inst{1}
        \and
        A. Matter\inst{1}
        \and
        A. Meilland\inst{1}
        \and
        F. Millour\inst{1}
        \and
        N. Nesvadba\inst{1}
        \and
        T. T. Shimizu\inst{7}
        \and
        E. Sturm\inst{7}
        \and
        G. Weigelt\inst{8}
        }

   \institute{Universit\'e C\^ote d'Azur, Observatoire de la C\^ote d'Azur, CNRS, Laboratoire Lagrange, Boulevard de l'Observatoire, CS 34229, 06304 Nice Cedex 4, France\\
              \email{jleftley@oca.eu}
    \and
    LESIA, Observatoire de Paris, Universit\'e PSL, CNRS, Sorbonne Universit\'e, Universit\'e de Paris Cit\'e, 5 place Jules Janssen, 92190 Meudon
    \and
    School of Physics \& Astronomy, University of Southampton, University Road, Southampton SO17 1BJ, UK
    \and
    Leiden Observatory, Leiden University, Niels Bohrweg 2, NL-2333 CA Leiden, The Netherlands
    \and
    MPI for Astronomy, Königstuhl 17, 69117 Heidelberg
    \and
    Univ. Grenoble Alpes, CNRS, IPAG, 38000, Grenoble, France
    \and
    Max-Planck-Institut für extraterrestrische Physik, Postfach 1312, 85741 Garching, Germany
    \and
    Max-Planck-Institut f\"ur Radioastronomie, Auf dem H\"ugel 69, D-53121 Bonn, Germany
             }

\date{Received date /
Accepted date }

\abstract{The Very Large Telescope Interferometer (VLTI) has been providing breakthrough images of the dust in the central parsecs of Active Galactic Nuclei (AGN), thought to be a key component of the AGN unification scheme and AGN host galaxy interaction. In single infrared bands, these images can have multiple interpretations some of which could challenge the unification scheme. This is the case for the archetypal type 2 AGN of NGC\,1068. The degeneracy is reduced by multi-band temperature maps which are hindered by ambiguity in alignment between different single band images. 
}
{To solve this problem by creating a chromatic model capable of simultaneously explaining the VLTI GRAVITY+MATISSE $2\,\mu$m$-13\,\mu$m observations of the AGN hosted by NGC\,1068.}{We employ a simple disk and wind geometry populated with spherical black body emitters and dust obscuration to create a versatile multi-wavelength modelling method for chromatic IR interferometric data of dusty objects.}{This simple geometry is capable of reproducing the spectro-interferometric data of NGC\,1068 from K$-$N-band, explains the complex single band images with obscuration and inclination effects, and solves the alignment problem between bands. {We find that the resulting model disk and wind geometry is consistent with previous studies of comparable and larger scales. For example, compared to molecular gas emission, our model wind position angle (PA) of $22^3_2{}^\circ$ is close to the mas scale outflowing CO(6--5) PA of $\sim33^\circ$ seen with {ALMA}. The equivalent $90^\circ$ offset model disk PA is also consistent with the CO(6--5) disk axis of $112^\circ$ as well as the mas scale disk axis from CO(2--1), CO(3--2), and HCO$^+$(4--3) of $115\pm5^\circ$}. Furthermore, the resulting model images visually resemble the multiple achromatic image reconstructions of the same data when evaluated at the same wavelengths. {We conclude that the IR emitting structure surrounding the AGN within NGC\,1068 can indeed be explained by the clumpy disk+wind iteration of the AGN unification scheme.} Within the scheme, we find that it is best explained as a type 2 AGN and the obscuring dust chemistry can be explained by a mix of olivine silicates and $16\pm1\%$ amorphous carbon.}{}

\keywords{Galaxies: Seyfert -- Galaxies: nuclei -- Galaxies: individual: NGC\,1068 -- Infrared: galaxies -- Techniques: interferometric}

\maketitle



\section{Introduction}

It is widely accepted that the spectral differences between type 1 (Sy1) and type 2 (Sy2) Seyfert active galactic nuclei (AGN), can be attributed to an obscuring medium that blocks an observers line of sight (LOS) to the central engine and broad line region (BLR) in Sy2s but not in Sy1s \citep{antonucci_unified_1993}. The obscurer is generally thought to be a dusty medium in the equatorial direction, such as a thick disk, that obscures the central region when the AGN is viewed edge on but not face on. The medium is also thought to be clumpy in order to both explain how a thick disk can be vertically supported \citep{krolik_molecular_1988} and observed short timescale variations of obscuration in X-rays \citep{risaliti_rapid_2005}. The medium then also acts as a material reservoir to fuel the central engine and emission regions \citep{honig_redefining_2019,hickox_obscured_2018}. Given the tremendous amounts of energy released by the central engine, some of the dusty material could be lifted out of the disk and into a dusty outflow through radiation pressure, opening up a pathway for AGN feedback to the host galaxy \citep[e.g.,][]{williamson_3d_2019,wada_radiation-driven_2012}. Therefore, understanding this dusty structure is crucial to understanding AGN inner dynamics and AGN - host galaxy interactions.

First generation long baseline IR interferometry offered a powerful new tool to AGN science by resolving the expected spacial scales of the obscuring dusty material at the wavelengths that it thermally emits. It provided strong evidence for polar dust structures, associated to a dusty outflow \citep{lopez-gonzaga_mid-infrared_2016,honig_dust_2013,burtscher_diversity_2013}, in the mid-IR (N-Band, $8-13\,\mu$m) by MIDI \citep[Mid-Infrared Interferometer,][]{leinert_midi_2003} and ring or disk-like structures \citep{weigelt_vlti/amber_2012,kishimoto_exploring_2009, kishimoto_innermost_2011} in the near-IR (K-band, $\sim 2.2 \mu$m) by AMBER \citep[Astronomical Multi-BEam combineR,][]{petrov_amber_2007} and Keck \citep{colavita_keck_2013}. The discovered geometry agreed with, and helped develop upon, our understanding of the dusty material leading to an iteration of the unification scheme that contains an obscuring equatorial dust-gas disk and an outflowing dusty wind \citep{honig_parsec-scale_2012}.

Second generation IR interferometers improved upon the first through increased \textit{uv} coverage per observation and improved phase information. AMBER in particular demonstrated the application of closure phases in IR interferometry, a lesson which GRAVITY \citep{gravity_collaboration_first_2017} and MATISSE \citep[Multi AperTure mid-Infrared SpectroScopic Experiment,][]{lopez_matisse_2022} have well received. 
Whereas first generation instruments such as MIDI and AMBER provided model derived geometry for AGN, MATISSE and GRAVITY can allow for image reconstructions \citep[e.g. for AGN:][]{isbell_dusty_2023,isbell_dusty_2022,gamez_rosas_thermal_2022,gravity_collaboration_central_2021,gravity_collaboration_image_2020}.

One such object is the Sy2 AGN, \object{NGC\,1068}. GRAVITY data of NGC\,1068 at $2.0-2.3\,\mu$m was reconstructed and a ring like structure was found \citep{gravity_collaboration_image_2020}. The structure had multiple interpretations in the work; however, the two favoured geometries were either: the ring is dust near the sublimation radius or the ring is coincidental and consists of clumpy dust on the illuminated back wall of the obscuring disk or inner outflow. Alternatively, a radiation transfer (RT) modelling attempt of the GRAVITY data suggested a thin ring perpendicular to the accretion plane \citep{vermot_inner_2021}, interpreted as possibly a tidal disruption event unrelated to the dust structure of the unification scheme.

\citet{gamez_rosas_thermal_2022} used MATISSE to observe and reconstruct images for NGC\,1068 from $3-13\,\mu$m and found similar structure to GRAVITY at the shortest wavelengths of MATISSE. However, the wider wavelength range provided the possibility to extract thermal information from the images derived at different wavelengths. By assuming the MATISSE and GRAVITY images at each wavelength were aligned by their matching spatial flux distributions and brightest spots, an SED was extracted for different apertures aligned to the observed dust structures, and a two black body and dust obscuration model was fit. The image alignment had the further constraint that the extracted SEDs must be continuous. This modelling suggested that the dust in the ring feature observed with GRAVITY was too cool ($\leq 600\,$K) to be sublimating (T$_\mathrm{sub} \approx 1500-1800\,$K). 
The caveat of this interpretation is that the inter-wavelength image position registration was assumed by geometry and SED instead of known. The absolute image position would require absolute phase information which is destroyed by the atmosphere. To know the relative alignment between bands instead would require inter-band differential phases which are not currently available. The uncertainty leaves room for ambiguity over which interpretation is correct. A different registration could give rise to different temperatures and different positions of the dust within the AGN.

Furthermore, \citet{gravity_collaboration_image_2020} and \citet{gamez_rosas_thermal_2022} provide multiple alignments between the image reconstructions and the H$_2$O maser emission of NGC\,1068, most recently published in \citet{gallimore_high-sensitivity_2023} who themselves also provide a additional alternate alignments. This makes it difficult to build a clear understanding of the coincidence between the maser and dust emission structures. It is clear that uncertainty in the image positions is a major limitation when studying structures at such high angular resolutions.

It is the aim of this work to provide an interpretation of the IR interferometric data that is free of a "visually based" image alignment though modelling the MATISSE and GRAVITY data simultaneously with a chromatic model compatible with the unified scheme of AGN. Specifically, we will design and use the model to:

\begin{itemize}
    \item attempt to explain the complex IR interferometric data and reconstructed images of the MATISSE \citep{gamez_rosas_thermal_2022} and GRAVITY \citep{gravity_collaboration_image_2020} observations,
    \item determine if the recovered model is compatible with the unification scheme of AGN,
    \item if so, determine if a face on, ring/disk dominated in KL-band, structure (Sy1) or edge on wind dominated structure (Sy2) is a better fit,
    \item use the best fit models to produce a model dependent image alignment and black hole location.
\end{itemize}

The derived geometry will be used 
for a full RT modelling of the dusty structure in NGC\,1068 in future work. The model we present will also 
be applied to a larger sample of AGN to create a more complete geometric description than was previously possible before second generation VLTI instrumentation.

\subsection{Radiative transfer models with chromatic spatially disperse information}

RT models have been used to interpret the SEDs of AGN successfully for many years \citep[e.g.,][]{gonzalez-martin_exploring_2019,stalevski_3d_2012,honig_dusty_2010,nenkova_agn_2008,nenkova_dust_2002}; they are flexible and robust tools for determining model dependent properties of dust in AGN. Combining the SED fitting process with simultaneous fitting of IR interferometric data has proved to be a very powerful tool with greater constraints on the dust geometry than the SED alone \citep{honig_dusty_2017}. SED modelling provides zero-order geometric information, the distance and distribution of the dust relative to the central engine can be derived but the absolute positions are ambiguous due to the unresolved nature of the information. By providing resolved information, distinctions can be made between otherwise well fitting models 
\citep[e.g.][]{isbell_dusty_2022,leftley_parsec-scale_2019,leftley_new_2018,honig_dusty_2017}. 

The results of models like CAT3D-WIND \citep{honig_dusty_2017} when simultaneously reproducing interferometric and SED data as well as the recent work of \citet{isbell_dusty_2022, gamez_rosas_thermal_2022} demonstrates 
the usefulness of combing spatially resolved observations with chromatic information when modelling dust structures. While SED modelling is well established, only after the first generation of VLTI IR interferometers were applied to AGN was it possible to do this on milliarcecond scales in the IR where the dust thermally emits \citep[e.g.,][]{stalevski_dissecting_2019}. With the second generation of instrumentation, it is now possible to apply these techniques to a much wider range of wavelengths and in greater detail due to improved \textit{uv} coverage per observation and phase information. We take inspiration from the simultaneous RT and interferometric modelling of previous works and create a chromatic model for AGN.

In Sect.\,\ref{s:Obs} we give the observations we model. In Sect.\,\ref{S:model} we describe in detail how the model is constructed and how we create brightness distributions per wavelengths. In Sect.\,\ref{s:Method} we detail how we compare our model to the data and determine the best set of parameters to reproduce the data. In Sect.\,\ref{s:Results} we provide the results of the modelling described previously with some discussion into what each set of results indicate about NGC\,1068 alone. {In Sect.\,\ref{sec:physdisc} we discuss the results in the context of the wider literature. In Sect.\,\ref{s:Discussion} we discuss the different fitting methods 
and prospects for future work.} Finally, in Sect.\,\ref{s:Conclusions} we give our final remarks and summarise the main points of the work.

\section{Observations}\label{s:Obs}

This work makes use of published observations of NGC\,1068 made with GRAVITY and MATISSE on the VLTI (priv. comm.)\footnote{Reduced MATISSE data can be found at \url{https://github.com/VioletaGamez/NGC1068_MATISSE}}. The detailed description of the observations and reduction for the GRAVITY and MATISSE data can be found in \citet{gravity_collaboration_image_2020} and \citet{gamez_rosas_thermal_2022}, respectively. The raw data can be found on the ESO archive under the following IDs: {MATISSE programme ID: 0103.C-0143, 0104.B-0322 GRAVITY programme ID: 0102.B-0667, 0102.C-0205 and 0102.C-0211.}

We use the same data as the previous works to remove the possibility of any differences from data handling. The only difference we make is that we do not use the Auxiliary Telescope (AT) data given in \citet{gravity_collaboration_image_2020} because combining Unit Telescope (UT) and AT data is a non-trivial problem due to the different FoVs between the telescopes.

\section{Model}\label{S:model}

To interpret the observational data, we create a semi-RT model of AGN dust morphology. The morphology we choose is based on the disk+wind interpretation of the unified model \citep[e.g.,][]{honig_parsec-scale_2012} due to its previous success in reproducing IR interferometric observations of AGN such as with CAT3D-WIND \citep{honig_dusty_2017}. CAT3D-WIND is a successful radiative transfer implementation of the disk+wind model. In their work they use CAT3D-WIND to model the IR SED of many AGN including those that CAT3D (a clumpy torus model) could not. It has also been used to model MIDI observations of AGN successfully \citep[e.g.]{leftley_new_2018}. More recently, it was used in one of the interpretations of the GRAVITY data we use in this work \citep[][model 4]{gravity_collaboration_image_2020}. Outside of CAT3D-WIND, the geometry has also been used to explain Circinus interferometric observations with RT models \citep{isbell_dusty_2022,stalevski_dissecting_2019}. Furthermore, the disk+wind model is backed up by hydrodynamic and numerical simulations that find a dusty wind can be easily formed from an accreting dust disk around a central engine \citep{williamson_radiation_2020, venanzi_role_2020, wada_multi-phase_2016,wada_obscuring_2015, wada_radiation-driven_2012}. IR radiation from the dust disk itself then pushes the wind vertically through radiation to form a hollow hyperbolic cone. Therefore, we decide to use this geometry in our model.

\subsection{The geometry}

We interpret the disk+wind model in terms of two components: a disk with an opening angle and inner radius, and two hyperbolic hollow cones that are perpendicular to the disk. We define our initial geometry 
such that $z$ is perpendicular to the plane of the disk, i.e. the polar axis of the AGN, and ($x,y$) are in the disk plane. For convenience, we also define $z$ to be North and $x$ to be East.

We define the disk to have an inner radius ($r_\mathrm{in}$) and an opening angle $2\phi$ where $\phi$ is a free variable (the half-opening angle); the disk opens from the inner radius. In a 2D plane, the surface of the disk can be defined as:

\begin{align}
    \begin{array}{lllll}
    z & = & \pm \left(\sqrt{x^2+y^2}-r_\mathrm{in}\right) \tan{\phi},&& \\[5pt]
     & = & \pm \left(r-r_\mathrm{in}\right) \tan{\phi},& \  \mathrm{for} \ & r>r_\mathrm{in} 
    \end{array}\label{eq:disk height}
\end{align}

where $r=\sqrt{x^2+y^2}$ and the half opening angle $\phi$ is defined from the disk plane ($x,y$). 

\begin{figure}[ht]
    \centering
    \includegraphics[width=\linewidth]{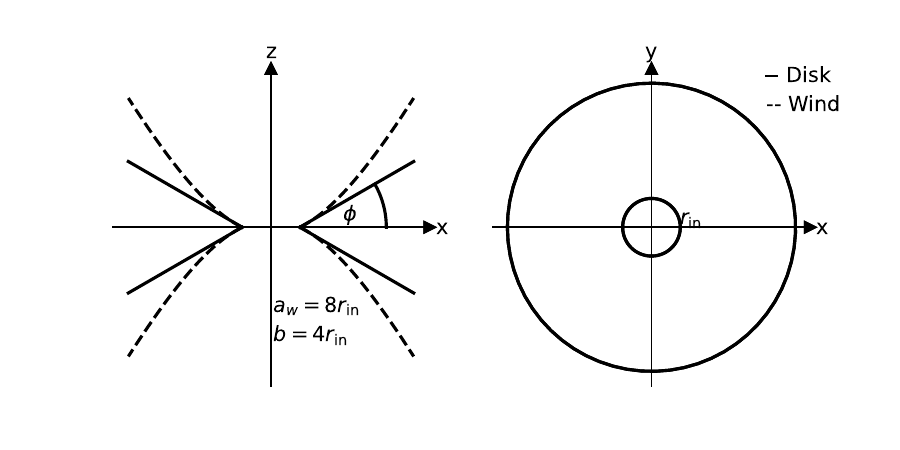}
    \caption{Schematic of the disk described in Equation\,\ref{eq:disk height} and wind defined in Equation\,\ref{eq:winddef}. We provide the example $a_w$ and $b$ used to make the wind schematic.}
    \label{fig:diskschem}
\end{figure}

We define the wind to be a hyperbolic cone described by:

\begin{align}
    \begin{array}{llll}
    z& = & \pm&a_w\left(\sqrt{\left(\frac{r}{b}\right)^2+1} - \sqrt{\left(\frac{r_\mathrm{in}}{b} \right)^2+1}\right),\\[10pt]
    &&& \  \mathrm{for} \ r>r_\mathrm{in}
    \end{array}\label{eq:winddef}
\end{align}
where $a_w$ and $b$ are fitted variables that define the hyperbolic cone. These two components are the base geometry of our model. A schematic of the disk and wind can be found in Fig.\,\ref{fig:diskschem}.

\subsection{Applying clumpiness}

We want our structure to be clumpy so for each component we distribute a number of clumps. We define two variables for the number of clumps. $N_p$ we give as the total number of clumps in the system and $f_w$ is the $\log_{10}$ fraction of $N_p$ that is attributed to the wind. Therefore, we can derive the number of clumps in the disk as  N$_\mathrm{d}=\left(1-10^{f_w}\right)N_p$ and the number of clumps in the wind as N$_\mathrm{w}=10^{f_w}N_p$.


To distribute clumps in the disk, we do so randomly with a probability based on the geometry in Equation\,\ref{eq:disk height}. The probability in each dimension is:
\begin{equation}
    P\left(x,y\right),P\left(z\right)  = \  0  \  \mathrm{for} \  r<r_\mathrm{in}, \nonumber\\
\end{equation}
\begin{align}
\begin{array}{llll}
    P\left(x,y\right)&=&A\exp{-\frac{x^2}{2\beta^2}}\exp{-\frac{y^2}{2\beta^2}},& \  \mathrm{for} \ r>r_\mathrm{in} \\
    P\left(z\right)&=&\frac{3}{\sqrt{2\pi}\left(r-r_\mathrm{in}\right)\tan{\phi}}\exp{-\frac{\left(3z\right)^2}{2\left(r-r_\mathrm{in}\right)^2\tan^2\phi}},\\
\end{array}\label{eq:diskprob}
\end{align}
where $A$ is the normalisation of the truncated Gaussian and $\beta$ is the standard deviation of the dust distribution in the plane of the disk. 
For $P\left(z\right)$ we have defined the disk height to be 3$\sigma$ of the normal distribution.

\begin{figure}[h]
    \centering
    \includegraphics[width=\linewidth]{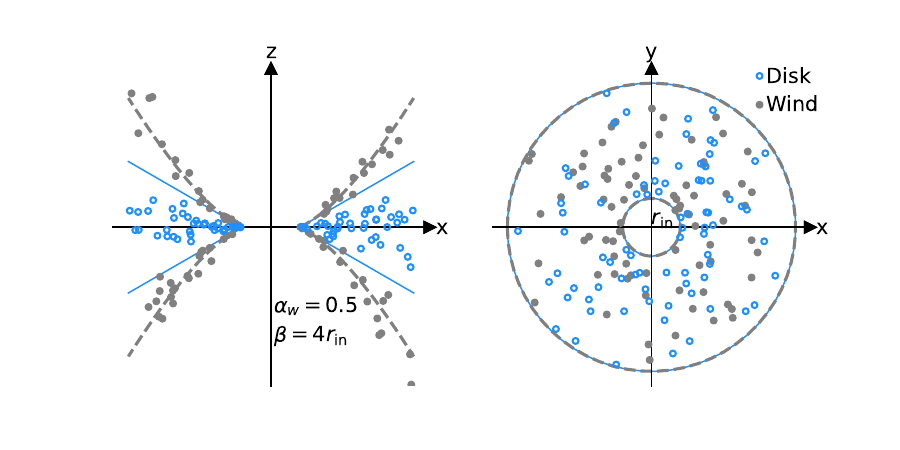}
    \caption{Schematic of the disk described in Equation\,\ref{eq:disk height} and wind defined in Equation\,\ref{eq:winddef} with clumps distributed with the probabilities described in Equations\,\ref{eq:diskprob} and \ref{eq:windprob}. We provide the example $\beta$ and $\alpha_w$ used to make the wind schematic.}
    \label{fig:dotschem}
\end{figure}

Along the wind, we distribute $N_w$ clouds along the hyperbolic cone in a similar manner to the disk. 
To give the wind a thickness, we randomly offset $a_w$ for each clump in the wind by a normal distribution with a standard deviation of $a_\mathrm{width}$, which can be a free parameter but is fixed in this work, and centred on $a_w$. 
The spatial probability distribution for clumps in the wind structure is therefore:
\begin{equation}
    P\left(x,y\right),P\left(z\right)  = \  0,  \  \mathrm{for} \  r<r_\mathrm{in} \nonumber\\
\end{equation}
\begin{align}
\begin{array}{lllr}
    P\left(x,y\right)
    & \propto & r^{-\alpha_w}& \mathrm{for} \ r_\mathrm{in}<r<r_\mathrm{out},\\
    P\left(a_w\right)& \propto &\exp{-\frac{a_w^2}{2a_\mathrm{width}^2}},
\end{array}\label{eq:windprob}
\end{align}
where $\alpha_w$ is a power law for the distribution of clumps in the wind in the ($x,y$) plane, and $r_\mathrm{out}$ is the outer radius defined as four times larger than the image field of view (FOV) of the longest wavelength being modelled. $r_\mathrm{out}$ is arbitrarily large and defined to save computation time by excluding points that fall too far away from the centre to contribute to the total AGN IR flux. When computing $z$ from Equation\,\ref{eq:winddef} for a clump, each point has an equal probability of being positive or negative to create a biconical outflow. An example schematic can be seen in Fig.\,\ref{fig:dotschem}.

We then incline the system towards the viewer (about the x-axis) by $inc$ and rotate in the image plane by $ang$ (about the y-axis). {$ang$ is equivalent to the on-sky from North through East Position Angle (PA) of the polar axis which is the axis perpendicular to the disk. In an aligned system, this is the direction of the jet.}

\subsection{Defining a clump}

In the previous section, we described how we distribute clumps in the geometry. Here we define what a clump is. We define all clumps as identical spheres of radius $r_c$ where the only unique properties of an individual clump is the position and temperature.

Each clump is given two temperatures, this is because AGN are centrally powered. If a clump is optically thick, then the side facing the central engine is directly heated to $T_\mathrm{p}$ while the other side is shielded and therefore at a lower temperature $T_b$. For the directly heated side of the clump, we set the temperature as a broken power law based on distance from the central engine. We set the power and break separately for each component. For the wind, we set
\begin{align}
    T_\mathrm{p}&=T_0r_\mathrm{in}^\alpha R^{-\alpha},\ &\mathrm{for}\ R<r_\mathrm{in} \alpha_r \nonumber\\
    &=T_0r_\mathrm{in}^{\alpha+\alpha_\mathrm{off}}\alpha_r^{\alpha_\mathrm{off}}R^{-\left(\alpha+\alpha_\mathrm{off}\right)},\ &\mathrm{for}\ R>r_\mathrm{in} \alpha_r
\end{align}
 where $R=\sqrt{x^2+y^2+z^2}$, $\alpha$ is the power before the break, $\alpha_\mathrm{off}$ is the additional power after the break, and $\alpha_r$ is the break radius in units of $r_\mathrm{in}$. In the disk we set 
 \begin{align}
    T_\mathrm{p}&=T_0r_\mathrm{in}^\gamma R^{-\gamma},&\ \mathrm{for}\ R<r_\mathrm{in} \gamma_r\nonumber\\
    &=T_0r_\mathrm{in}^{\gamma+\gamma_\mathrm{off}}\gamma_r^{\gamma_\mathrm{off}}R^{-\left(\gamma+\gamma_\mathrm{off}\right)},&\ \mathrm{for}\ R>r_\mathrm{in} \gamma_r
\end{align}
 where $\gamma$ is the power before the break, $\gamma_\mathrm{off}$ is the additional power after the break, and $\gamma_r$ is the break radius in units of $r_\mathrm{in}$. In this work, we do not fit the SED of NGC\,1068 which makes the result insensitive to the absolute temperature; only the relative temperature is required. Therefore, we set the power law normalisation $T_0$ to $1800\,$K at $r_\mathrm{in}$.

 The second temperature $T_b$ is also set separately for the disk an the wind. We define this temperature as a variable constant for each component. For the disk we define the variable as $T_{bd}$ and for the wind we define it as $T_{bw}$.

 \subsubsection{Optical depth of a clump}

We assume a clump is uniform in density and we can define the optical depth ($\tau$) along the LOS for a clump as a function of $\left(x,z\right)$. The equation for the optical depth is 
\begin{equation}
    \tau_c\left(r'_c,\lambda\right) = \left(2\tau_{c0}\left(\lambda\right)\sqrt{r_c^2-{r'_c}^2}\right)\mathrm{\ for\ }r'_c \leq r_c,
\end{equation}
where $\tau_{c0}\left(\lambda\right)$ is the wavelength dependent attenuation coefficient of the clump, $r'_c=\sqrt{\left(x-x_c\right)^2 + \left(z-z_c\right)^c}$, and ($x_c,z_c$) is the ($x,z$) location of the clump.

\subsection{Defining emission}

Now we have a clump temperature and optical depth, we can define the emission. When defining emission for a clump, we evaluate per pixel of the image. A pixel has a coordinate $\left(i,j\right)$ where $i$ and $j$ are $x$ and $y$ scaled by the pixel size in mas. We assume that a clump emits thermally as two black bodies such as $\left(1-\exp{\left(-\tau_c\left(r'_c\left(i,j\right),\lambda\right)\right)}\right)B_\nu\left(\nu,T_\mathrm{p}\right)$ and $\left(1-\exp{\left(-\tau_c\left(r'_c\left(i,j\right),\lambda\right)\right)}\right)B_\nu\left(\nu,T_\mathrm{b}\right)$ where $B_\nu$ is Planck's law. For $T_\mathrm{p}$, we calculate the total flux from this temperature for a clump as $F_\mathrm{p}\left(\lambda\right)\propto\left(1-\exp{\left(-\tau_c\left(r'_c,\lambda\right)\right)}\right)B_\nu\left(\nu,T_\mathrm{p}\right)$. 
However, we defined this temperature component as the side that faces the central engine and at some viewing angles the clump will self-obscure if optically thick. We approximate this phase angle effect (sometimes referred to as moon phase effect) by applying a factor to the flux of a clump based on its position relative to the centre of the model and the viewer: 
\begin{align}
    \theta_l &= \arctan \frac{z_c}{-y_c}, \nonumber\\
    \theta_m &= \arctan \frac{-y_c}{x_c}.
\end{align}
Therefore, the flux factor can be defined as $S_\mathrm{moon} = 0.5\left(\cos{\theta_l}\sin{\theta_m}+1\right)$. This is the same treatment that is used in some BLR cloud models \citep{rakshit_differential_2015}. The effect scales with optical thickness, so we apply a factor of $s_c\left(1-\exp{\left(-1.5r_c\sigma_c \left(\lambda\right)\right)}\right)$ to $S_\mathrm{moon}$, where $s_c$ is a free variable scaling, which scales the influence of the phase angle effect ($P_a$) with optical thickness. So $F_\mathrm{p}$ can be fully defined as:
\begin{align}
\begin{array}{lll}
    S_\mathrm{moon}& = &0.5\left(\cos{\theta_l}\sin{\theta_m}+1\right),\\[1.5pt]
    P_a\left(\lambda\right)&=&s_cS_\mathrm{moon}\left(1-\exp{\left(-\tau_c\left(1.5r_c,\lambda\right)\right)}\right),\\[1.5pt]
    F_\mathrm{p}\left(\lambda\right)& = & P_a\left(\lambda\right)\left(1-\exp{\left(-\tau_c\left(i,j\right)\right)}\right)B_\nu\left(\nu,T_\mathrm{p}\right).
    \end{array}
\end{align}

For the flux derived from the base temperature, we do not need to apply this phase angle effect. Therefore, the total flux contribution from the cooler side is
\begin{align}
\begin{array}{lll}
    F_\mathrm{cool}\left(\lambda\right)& = & \left(1-\exp{\left(-\tau_c\left(i,j\right)\right)}\right)B_\nu\left(\nu,T_\mathrm{b}\right).
\end{array}
\end{align}

Note that the flux is per Steradian. The phase angle effect accounts for the differences in emitting area for the two flux components so they can be additively combined before accounting for surface area. We account for this area when creating the image.

\subsection{Applying obscuration}\label{s:applying_obsc}

When creating an image, we apply two forms of obscuration. The first is from the clumps which we have already defined. A clump from the viewers perspective attenuates the flux behind it by $1-\exp\left({-\tau_c\left(r'_c,\lambda\right)}\right)$. The second form of obscuration is from a smooth dust structure ({the blue solid lines in Fig.\,\ref{fig:dotschem})}.

The smooth dust structure we define as smooth medium only present in the disk of Equation\,\ref{eq:disk height} that only acts in obscuration. This obscuration can be applied analytically. For simplicity, we define the obscuration in the plane of the disk such that $\left(x_r,y_r,z_r\right)$ are the same as the $\left(x,y,z\right)$ in Fig.\,\ref{fig:diskschem} and $r_r=\sqrt{x_r^2+y_r^2}$, i.e. before the geometry was rotated by $inc$ and $ang$. We choose the optical depth of the smooth structure to decrease in the $\left(x_r,y_r\right)$ plane and be uniform in $z_r$ such that
\begin{align}
\begin{array}{lll}
    \tau\left(x_r,y_r,z_r,\lambda\right)&\propto&\tau_0\left(\lambda\right)\exp{-\frac{x_r^2}{2\beta^2}}\exp{-\frac{y_r^2}{2\beta^2}},\\[4pt]&& \  \mathrm{for} \ r_r>r_\mathrm{in} \\[1.5pt]
    && \  \mathrm{and} \ \left(r_r-r_\mathrm{in}\right) \tan{-\phi}<z_r<\left(r_r-r_\mathrm{in}\right) \tan{\phi}\\[4pt]
    \tau\left(x_r,y_r,z_r,\lambda\right)&=&0, \  \mathrm{elsewhere}
\end{array}
\end{align}
to mimic the $\left(x_r,y_r\right)$ clump distribution in the same structure where $\tau_0\left(\lambda\right)$ is the attenuation coefficient at $r_r=0$ not accounting for $r_\mathrm{in}$. If we rotate the smooth obscuring geometry by applying the rotation matrix
\begin{align}
    \left[\genfrac{}{}{0pt}{0}{y}{z}\right]
    \ = \ 
     \left[\genfrac{}{}{0pt}{0}{\cos{\left(inc\right)} \ \ -\sin{\left(inc\right)}}{\sin{\left(inc\right)} \ \ \cos{\left(inc\right)}}\right]
    \left[\genfrac{}{}{0pt}{0}{y_r}{z_r}\right]
\end{align}
the LOS from an emitter to the viewer is along $y$ only allowing for a 1D integration. If the LOS from an emitter enters the disk at $\left(x_1,y_1,z_1\right)$ and exits at $\left(x_1,y_2,z_1\right)$ then the obscuration from the smooth structure can be defined as
\begin{align}
\begin{array}{lll}
    \tau\left(inc,\lambda\right) &=& 
    \tau_0\left(\lambda\right)\sqrt{1+\tan^2\left(inc\right)}\exp{\left ( -\frac{x_1^2}{2\beta^2} \right )}\\&&\times\int_{y_1}^{y_2}\exp{\left (-\frac{y^2}{2\beta^2  } \right ) }  \  dy.
\end{array}
\end{align}
This can be redefined in terms of the error function 
\begin{align}
\begin{array}{lll}
    \tau &=& 
    \tau_0\left(\lambda\right)\sqrt{1+\tan^2\left(inc\right)}\exp{\left ( -\frac{x_1^2}{2\beta^2} \right )}\\&&\times \left[\sqrt{\frac{\pi}{2}} \beta \,erf{\left (\frac{y}{\sqrt{2}\beta } \right ) }\right]^{y_2}_{y_1},
\end{array}
\end{align}
where we use the standard definition of the error function:
\begin{equation}
    erf{\left(t\right)} = \frac{2}{\sqrt{\pi}} \int^z_0 \exp{\left( -t^2 \right)}\ dt.
\end{equation}
In the completely face on case where $inc=\pi/2$, $\tau$ is ill-defined by the given parameterisation. Instead, the limit as $inc\rightarrow\pi/2$ should become the uniform dust case. We do not expect to encounter this Blazar-like case so we do not include it.

{In reality, we do not make $\tau_0\left(\lambda\right)$ or $\tau_{c0}\left(\lambda\right)$ a model parameter, instead we define $\beta\tau_0\left(\lambda\right)=n_d\sigma\left(\lambda\right)$ and $\tau_{c0}\left(\lambda\right)=n_c\sigma\left(\lambda\right)$. $n_d$ and $n_c$ are the model parameters and are in units of g\,cm$^{-2}$ and g\,cm$^{-2}$\,mas$^{-1}$, respectively. $\sigma\left(\lambda\right)$ is the optical cross section per mass in cm$^{2}$\,g$^{-1}$ of the dusty material used. We use the same dust mix for the clumps and the smooth structure. $n_c$ and $n_d$ are density scaling factors.}

{For the dust we apply in this work, we choose a mix of amorphous carbon and olivine silicates \citep{min_modeling_2005} as was found in \citet{gamez_rosas_thermal_2022} where we allow the relative amount of carbon to olivines be a free parameter $p_g$ in the same manner as in their SED fitting.  $\sigma\left(\lambda\right)$ for the two dust species can be seen in Fig.\,\ref{fig:dustcomp}. }

\subsection{Model evaluation specifics}

We have fully defined our model geometry and described the characteristics of its emission and absorption. However, there are some important assumptions we make when comparing the model to data. Here we discuss how we evaluate the model.

\subsubsection{PSF effects}

There is one final flux scaling factor that we apply when determining fluxes that is due to the telescope PSF. The scaling factor is to account for the flux drop from the telescope PSF and is approximated by a 2D Gaussian on the plane of the sky with a FWHM of $\lambda/D$ centred on the model centre. {This reduces the flux of a point by the distance from the centre.} This approximation does not account for the effects from fibre injection in GRAVITY and the pinhole and pupil stops of MATISSE but it is sufficient for our image sizes. It is noteworthy that we include this factor to ensure any bright points far from the centre are treated properly but the effect is usually negligible due to the PSF of a UT. For a telescope diameter of 8\,m the FWHM at 2\,$\mu$m is $\sim$50\,mas which is 1.7 times the FOV at the same wavelength.

\subsubsection{Making an image}

Before comparing our model to data, we must create an image per wavelength. To make our image we take our defined model, with the analytical smooth obscuration applied, and determine the clump emission and attenuation for all clump occupied pixels in the image. We assume that the obscuration at the centre of a pixel is a good approximation of the average value in the pixel when evaluating the clumps emission and obscuration. For pixels that are only partially filled, we reduce the flux by the fraction of the pixel that is unfilled. To account for changes in pixel size between wavelengths, we normalise the clump flux such that
\begin{equation}
    \int^{r_c}_{-r_c}F_\mathrm{c,\lambda}dr'_c = F_\mathrm{p,\lambda}+F_\mathrm{cool,\lambda},
\end{equation}
in the case where $\tau\rightarrow\infty$. So for any given pixel we get
\begin{equation}
     F_{i,j,k}=F_{i,j,k-1}\exp{\left(-\tau_{c}\left(r'_{c\,ijk}\right)\right)} + f_{ijk},
\end{equation}
for pixel $ij$ and clump-in-pixel $k$ where $k$ ranges from 1 to the number of clumps in the pixel. Furthermore, $F_0=0$, $f_{ijk}$ is the flux of clump $k$ for pixel $ij$, and $r'_{c\,ijk}$ is the in-clump radius $r'_c$ for clump $k$ at pixel $ij$. Should a clump fall outside an image, the flux is not included within the image but it is accounted for.

\subsubsection{Over-resolved structure}\label{sec:over-resolved}

{When observing extended objects with interferometry, flux can exist outside the largest angular scale the interferometer probes but still within the single dish photometric aperture. We define this as over-resolved structure. 
In AGN, this can be from the dust itself or unrelated emission from the host galaxy. Instead of images, IR interferometry provides observable quantities that can be used to reconstruct the target source. The two we model are the squared visibility (V$^2$) and closure phase ($\Phi_c$).} We compare the model V$^2$ and $\Phi_c$, calculated from the model image using the python software package \textsc{galario}, to the observed V$^2$ and $\Phi_c$, respectively. Any {over-resolved} flux that falls outside the image does not contribute to $\Phi_c$ but it does contribute to V$^2$ as a {PA independent }scaling factor. V$^2$ is defined as the squared flux observed by the interferometer divided by the squared flux observed by a single telescope. If you assume all the flux observed by the telescope falls within the image {when it does not}, your single telescope model flux will be lower than it should be {and V$^2$ will be higher in turn}. 

Additional scaling to V$^2$ can arise from instrumental effects causing flux losses in the interferometer. Therefore, when modelling a single wavelength or a small band it is common to apply a scaling factor to the model V$^2$ such that V$^2 = \left(V_0\mathrm{V}_\mathrm{model}\right)^2$ \citep[e.g.][]{gamez_rosas_thermal_2022,gravity_collaboration_resolved_2020,leftley_new_2018}. This achromatic factor has been shown to be a good approximation for a single wavelength. However, it will fail to capture any chromatic over-resolved structure. Because our model is chromatic with multiple instruments, applying an appropriate scale factor is not trivial.

To simplify the problem, we make the assumption that the instrumental losses are insignificant or do not change significantly between GRAVITY and MATISSE; this is a relatively safe assumption because NGC\,1068 is bright for an AGN and the data for both instruments has been very well reduced by the respective teams. Furthermore, the greatest source of losses for AGN IR interferometry, particularly at shortest wavelengths, is thought to be the MACAO AO system \citep{leftley_resolving_2021}. The GRAVITY data was observed under excellent conditions \citep[optical coherence time of 7-13\,ms][]{gravity_collaboration_image_2020} and the MATISSE data was filtered to remove bad AO performance \citep{gamez_rosas_thermal_2022}. Therefore, the losses from AO should be minimal. For fainter AGN this may not be appropriate and will need to be readjusted. Therefore, we just account for sources of over-resolved emission. We define our over-resolved structure with the following three components:

\vspace{5pt}
\noindent\textit{Over-resolved model flux}
\vspace{3pt}

Some of this over-resolved flux is inherently accounted for in our model. We sum the flux of all clumps that fall outside an image which we use to scale the model V$^2$. For computational efficiency, we only apply clump obscuration for clumps inside the image at that wavelength. We assume that clumps that fall outside the image are too sparsely distributed to significantly overlap each other and therefore we can sum just the emission from these clumps after only smooth obscuration. If you sum the flux inside and outside the image, you receive the total flux of the model AGN. {This is the direct contribution of the model dust to the over-resolved structure, the following components are not necessarily related to our dust structure but still contribute flux in the IR.}

\vspace{5pt}
\noindent\textit{A hot component}
\vspace{3pt}

We find that we do still require an additional factor to explain the shortest and longest wavelengths simultaneously. We apply a hot component to the over-resolved flux. Unlike the other two over-resolved components, we only apply this one when performing multi-band fitting.

\begin{figure}
    \centering
    \includegraphics[width=0.48\textwidth]{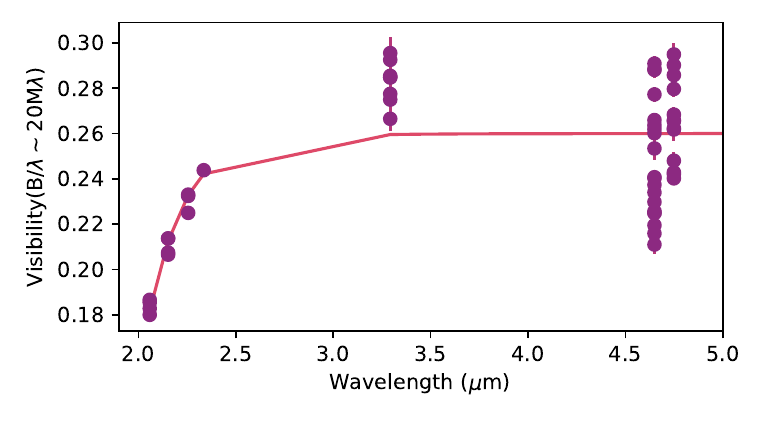}
    \caption{Visibility of NGC\,1068 between 16\,M$\lambda$ and 20\,M$\lambda$ compared to wavelength in purple. Overplotted is a best fit model of an over- and under-resolved black-body model in red.}
    \label{fig:bbshort}
\end{figure}

If we assume the simplest case that the structure responsible for the shortest baseline ($\sim40$\,m or $\sim18$\,M$\lambda$) V$^2$ at 2.05\,$\mu$m is the same for the equivalent baseline at 3.29\,$\mu$m then we find V$^2$ doubles between the two wavelengths (V$^2$ increases from 0.05 to 0.1 between 2.05\,$\mu$m and 3.29\,$\mu$m, see Fig.\,\ref{fig:bbshort}). We fit a simplistic model of $V\left(\lambda\right)=V_b \ \frac{\left(1-frac\right) \ BB_\mathrm{under}\left(\lambda,T_1\right)}{frac\ BB_\mathrm{over}\left(\lambda,T_2\right) +\left(1-frac\right)\ BB_\mathrm{under}\left(\lambda,T_1\right)}$, where $BB$ is a black body of temperature $T$ normalised to 1 at 2$\,\mu$m for an under- and over-resolved component; $frac$ is the relative flux of the under-resolved component at 2$\,\mu$m; and $V_b$ is an achromatic visibility, to visibilities between 16\,M$\lambda$ and 20\,M$\lambda$ from $\lambda=2\,\mu$m to $\lambda=5\,\mu$m. We find that the change can be explained by $T_2\sim20000\,$K,  $T_1\sim440\,$K, $frac\sim0.3$, and $V_b\sim0.26$ as seen in Fig.\,\ref{fig:bbshort} as the red line. The fit tells us that the over-resolved component can be explained by a hot black-body of $20000\,$K responsible for 30\% of the flux at $\lambda=2\,\mu$m. This does not rule out that the component could be colder with some change in visibility originating from a more complex structure than the unchanging single temperature geometry source assumed in the quick fit.

As such, we introduce an over-resolved hot black body of 20000\,K to our model when fitting multiple bands. We scale the flux from this black body based on the AGN flux such that $F_{BBh}\left(\lambda\right)=BB_\mathrm{hot}F_{AGN}\left(\lambda_{min}\right)B_\lambda\left(\lambda,20000\,\mathrm{K}\right)/B_\lambda\left(\lambda_{min},20000\,\mathrm{K}\right)$. $\lambda_{min}$ is the shortest wavelength modelled. We fix the temperature because we find it to be otherwise poorly constrained.

{There are multiple possibilities for the physical origin of this component. Under the assumption that the component is the Raleigh-Jeans tail of a hot black body, while we cannot constrain the exact temperature, we can conclude that it is too hot to be thermal emission from dust and not a missing component of our AGN thermally emitting dust model; i.e. it behaves as a power-law against wavelength with a power of -2 in our covered wavelength range. 
Previous work using Keck by \citet{weinberger_diffraction-limited_1999} finds that 50\% of the K-band nuclear (sub 0".1) flux comes from an extended region not attributed to direct heating by the AGN but instead scattered nuclear emission. \citet{weigelt_diffraction-limited_2004} followed this with K-band speckle interferometry which found an elongated Gaussian structure with a lower limit FWHM of $26\times58$\,mas responsible for 54$\pm5$\% of the nuclear K-band flux with the remaining from a compact nuclear source. The extended component in their work is proposed to be a mix of scattered light from dust in an outflow and thermal emission. Such extended scattered light from the accretion disk would be sufficient to explain the over-resolved emission at 2\,$\mu$m. Alternatively, star formation could reach sufficient temperatures of $\sim10000$\,K and \citet{lopez-rodriguez_near-infrared_2015} estimate that $45\pm16\,\%$ of the polarised K-band flux in the central 0".5 arises from stellar contamination.}

{Under the assumption that the component's emission is non-thermal, we find that while a power law with wavelength is just as good a description of the data, the power is completely degenerate with the temperature of the unresolved component with an upper limit on the power of 0. This means the non-thermal radio emission, with a power of $\lambda^{0.3}$ is also a possible contributor \citep{wittkowski_diffraction-limited_1998}. In reality, the over-resolved structure is likely to be from a combination of sources.}

{In future work, the inclusion of the SED when fitting or further investigation into the central region of NGC\,1068 with high angular resolution imaging may provide better constraints on the temperature and physical origin of this component. We note that, because we do not model absolute fluxes, the exact spectral shape of the component beyond being more influential in K- than L-Band will have minimal impact on our model result.}

\vspace{5pt}
\noindent\textit{An achromatic component}
\vspace{3pt}

{We additionally introduce an achromatic $V_0$ term. We find we require some scaling at longer wavelengths that is not covered by the hot black body. We apply this scaling as a wavelength independent multiplicative factor to the visibility at all wavelengths. This could be replaced by a second cooler over-resolved black body in future versions of the model. If physical and not instrumental, the origin of this factor could be from an extended larger smooth structure such as the northern component 3 revealed with MIDI data by \citet{lopez-gonzaga_revealing_2014}. The component from their work is suggested to be a cool 300\,K optically thin thermally emitting component but it is also mentioned that it could be a warmer ($\sim700-800\,K$) not heated by the central engine or instead part of the radio emission. The last possibility is not preferred because their component 3 is not well-aligned with radio emission. Additionally, \citet{rouan_hot_2004} find 475\,K dust at 450\,mas from the centre from high angular resolution observations with NACO. The spectral shape of this component is therefore unclear and the visibility is sufficiently approximated by the simpler achromatic term. With the inclusion of AT data or the SED, it may be possible to constrain this component further. This will not have significant impact in this work and can be investigated in more detail in future work.}

\subsection{Defining terms}

When performing the modelling, our choice of parameters differs in some cases from our definitions used to describe the model above. Here we summarise the parameters we use when modelling and how they relate to the model described above.




\begin{figure}
    \centering
    \includegraphics[width=0.48\textwidth]{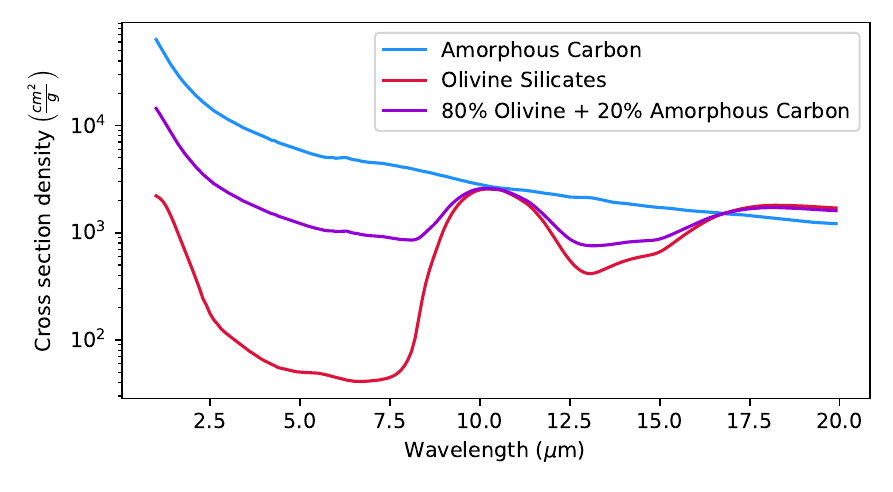}
    \caption{Optical cross section per gram of the dust materials used in this work as well as a 20\% amorphous carbon and 80\% olivine mix ($p_g=0.2$).}
    \label{fig:dustcomp}
\end{figure}

\subsubsection{Model parameter summary}

To summarise, the model has 27 free parameters, 11 of which are geometric parameters and 13 of which relate to the point source fluxes. We also have 1 additional fixed geometric parameter.{ The parameters, with descriptions, can be found in Tab.\,\ref{tab:bounds}.} 
Not included in the table are $\ln f_v$ and $\ln f_c$. $f$ is the fractional amount for which the variance is underestimated by the likelihood function if the errors were assumed correct \citep{foreman-mackey_emcee:_2013} and is defined separately for closure phase and $V^2$ as $f_c$ and $f_v$, respectively. We define the likelihood for one observable as

\begin{align}\label{eq:likelyhood}
         2\ln{\left(L\right)}&=&-\sum_m^N\left[\frac{\left(y\left(x_m\right)-\mathrm{model}\left(x_m,\alpha\right)\right)^2}{\sigma_m^2+f^2\mathrm{model}\left(x_m,\alpha\right)^2}+\right.\nonumber\\
         &&\left.\ln\left(2\pi\left(\sigma_m^2+f^2\mathrm{model}\left(x_m,\alpha\right)^2\right)\right)\vphantom{\frac{y^2}{y^2}}\right].
\end{align}

\section{Method}\label{s:Method}

We use the model described in Sect.\,\ref{S:model} with the \textsc{python} Markov Chain Monte Carlo (MCMC) fitting code \textsc{emcee} \citep{foreman-mackey_emcee:_2013}. We make use of the \textsc{galario} \citep{tazzari_galario_2018} and \textsc{astropy} \citep{astropy_collaboration_astropy:_2013} libraries. While the distribution of points is different, the model fitting process is the same as we used in the point source model of \citet{leftley_resolving_2021} and \citet{gamez_rosas_thermal_2022}. However, we do not penalise the addition of points because, unlike in the previous works, additional points do not require additional parameters.

Our model is polychromatic and can be fit to the entire wavelength range of GRAVITY and MATISSE simultaneously. However, the wider the wavelength range, the more likely we are to encounter complex chromatic structures not covered by our simple model. Therefore, we run the model in two different modes, an individual band fit and a full polychromatic fit. We can then compare the results from each to infer the presence of other structures if such structures are present.

Because we calculate an image cube of models when fitting to compare with the observables at each wavelength, it is necessary to bin the observational data in the wavelength direction when fitting due to computational resource limitations. However, too large a binning removes the chromatic information contained within a single band. Therefore, we decide on the following binning in each band: from 3.2$\,\mu$m $-$ 3.9$\,\mu$m in 4 bins for L-band, 4.6$\,\mu$m $-$ 4.8$\,\mu$m in 2 bins for M-band, and 8.40$\,\mu$m $-$ 9.0$\,\mu$m in 1 bin, 9.40$\,\mu$m $-$ 11$\,\mu$m in 3 bins, and 11.4$\,\mu$m $-$ 12.7$\,\mu$m in 2 bins for N-band. K-band is unbinned due to the rapid change in structure in the band coupled with the low spectral resolution from the fringe tracker data. We do not use the shortest wavelength bin due to the known contamination by the metrology laser \citep{gravity_collaboration_image_2020} so we have 4 wavelength points in K-band. We use the same binning in both the band-by-band and polychromatic modelling.

\subsection{band-by-band} \label{sec:MethodBbB}

At the shortest wavelengths, we are sensitive to the relative position of individual points, i.e. we resolve the non-uniform distribution of dust in the AGN. Due to the random nature of the point distribution this means we have a dependence on the random seed. We expect this to be particularly influential in the edge-on cases because the inner radius ring-like distribution dominates in the face-on case which becomes smooth at smaller $N_p$ than the more disperse wind distribution. To evaluate the effect per band, we perform the fit multiple times for different seeds in the band-by-band fit. From this we can determine the specific best fit, the median best fit, and the uncertainty in the model and the parameters introduced by randomness. We also perform the fit twice, once imposing that the central region be obscured ($inc<\phi$) and once imposing an unobscured central region labelled Sy2 case and Sy1 case, respectively. When we do not impose this, we find that inclination is linearly correlated with wavelength which is potentially unphysical.

We define two sets of boundaries given in Table\,\ref{tab:bounds} and fit the model at the wavelength ranges given in Sect.\,\ref{s:Method}. We randomly select a list of unique random seeds and run an MCMC fit for each band for each seed. The initial distribution of walkers is distributed uniformly over the initial range given in Table\,\ref{tab:bounds} which are not the same as the boundaries given for the fit. The starting ranges were selected to cover the ranges to which the model is most sensitive, less computationally expensive, or most physically reasonable from hydrodynamic simulations \citep{williamson_3d_2019,williamson_radiation_2020}.

\begin{table*}
    \caption{Model Parameters, Bounds, and Initial Values}
    \centering
    \begin{tabular*}{\textwidth}{@{\extracolsep{\fill}} ccccccc}
        \hline
        \multicolumn{1}{c}{Parameter}&Units&\multicolumn{1}{c}{Description}&\multicolumn{2}{c}{Bounds}&\multicolumn{2}{c}{Initial}   \\
        &&&Lower&Upper&Lower&Upper\\ 
        \hline
        $\alpha$&-&Wind temperature power&1E-5&5&0.1&0.8\\
        $\alpha_\mathrm{off}$&-&Wind temperature additional post break power&0&2&0&1.5\\
        $\beta$&mas&Standard deviation of disk dust&1E-6&400&0.1&300\\
        $\phi$&deg&Half-opening angle&0&45&5&40\\
        $b$&-&Hyperbolic shape parameter&1&4&0.1&2\\
        $r_\mathrm{in}$&mas&Sublimation radius&2&5&3.4&3.6\\
        $\gamma_r$&$r_\mathrm{in}$&Disk power break radius&1&5&2&4.6\\
        $\alpha_r$&$r_\mathrm{in}$&Wind power break radius&1&5&2&4.6\\
        $a_w$&-&Hyperbolic shape parameter&0.1&4&0.1&2\\
        {$a_\mathrm{width}$}$^+$&-&Standard deviation for normal distribution about $a_w$&0.05&-&-&-\\
        $n_c$&$\log_{10}$(g\,cm$^{-2}$\,mas$^{-1}$) &Clump dust density&-7&0&-3.5&-2.5\\
        $n_d$&$\log_{10}$(g\,cm$^{-2}$ $\beta$)& Normalised smooth dust density at $r=0$&-5.5&-2&-4.5&-2.5\\
        $\gamma$&-&Disk temperature power&1E-5&5&0.1&5\\
        $\gamma_\mathrm{off}$&-&Disk temperature additional post break power&1E-5&5&0&1.5\\
        $inc$&deg&The inclination&0&90&$0||45$*&$39||89$*\\
        $ang$&deg&The PA from North of the polar axis&0&360&5&45\\
        $p_g$&-&Percentage of dust that is amorphous carbon&0&1&0&0.3\\
        ${f_v}^\dag$&-&V$^2$ error underestimation&1E-15&1&1E-10&1E-9\\
        ${f_c}^\dag$&-&$\Phi_c$ error underestimation&1E-15&1&1E-10&1E-9\\
        $N_p$&-&Number of points&90&1E5&1E2&2E4\\
        ${f_w}^\dag$&-&Log$_{10}$ fraction of points in wind&1E-2&1&1E-2&1\\
        $\alpha_w$&-&Radial power for distribution of wind clumps&0&3&0&3\\
        $s_c$&-&Moon phase effect relative intensity&0.1&1.1&0.5&1.1\\
        $r_c$&mas& Clump radius&0&5&0&0.4\\
        $BB_\mathrm{hot}^\ddag$&$F_\mathrm{AGN}\left(\lambda_\mathrm{min}\right)$&Flux of hot component at shortest wavelength&0&2&-&-\\
        $V_0$&-&Achromatic over-resolved component&0&1&0.4&1.0\\
        $T_\mathrm{bw}$&K&Wind clump base temperature&0&400&0.5&300\\
        $T_\mathrm{bd}$&K&Disk clump base temperature&0&400&0.5&300
        
    \end{tabular*}
    \begin{tabular}{@{\extracolsep{\fill}} c}
        \\
        \hline
        Additional rules\\
        $\phi<90-\tan^{-1}\left(b/a_w\right)$\\
        $\phi>inc || \phi<inc$*
    \end{tabular}
    
    \tablefoot{\tablefootmark{$+$}{Fixed parameter,} \tablefootmark{$\dag$}{fit in log$_{10}$,} \tablefootmark{$\ddag$}{not used in band-by-band fit,} \tablefootmark{*}{for the edge-on band-by-band and face-on band-by-band fits, respectively.}}
    
    \label{tab:bounds}
\end{table*}

We select the best result per wavelength from all samples of all seeds through Equation\,\ref{eq:likelyhood}. This becomes the starting point of a subsequent run of the same seeds with walkers distributed normally around the result with a standard deviation of 5\%. Once fitted, when calculating the median fit model observables (V$^2$ and $\Phi_c$) and their uncertainties, we randomly sample the pool of walkers 1000 times with replacement and determine the observables for each choice. We then take the 16$^\mathrm{th}$, 50$^\mathrm{th}$, and 84$^\mathrm{th}$ percentiles of the 1000 observables to find the error and median. The error on the parameters themselves are calculated for each seed from the 16$^\mathrm{th}$ and 84$^\mathrm{th}$ percentiles with the given value being the 50$^\mathrm{th}$ percentile not including burn-in samples. When giving average values of parameters all seeds together, we combine the sampler pools from all seeds before deriving the median and errors.

\subsection{Polychromatic fit}\label{S:M:poly}

We proceed to fit all provided wavelengths simultaneously. We use the parameters found in the band-by-band modelling as prior information to set the initial walker distribution. Furthermore, we only fit for 1 seed. For the majority of the parameters, we used the results from the M-band because this lays in the centre of our coverage. However, we selected $r_c$, $s_c$, and $p_g$ from the K-band as they are most sensitive to the shortest wavelengths. Furthermore, we select $n_d$, $n_c$, $\phi$, and $\beta$ from the N-band due to the silicate feature and larger FoV. While this was the procedure we decided upon, upon fitting we found the increase in data coverage made the final model largely insensitive to the choice of initial walker position. Finally, similar to the band-by-band fit, we perform a second longer modelling run from the best result. The results and statistics we present are from the second run.

\section{Results}\label{s:Results}

In the initial modelling of the band-by-band fit, we find the Sy2 case to be a better fit that the Sy1 case by likelihood. Because the Sy2 case was preferred, we perform the final fit on this case only. As described in Sect.\,\ref{sec:MethodBbB}, we initiated the final fit for each seed from the best parameters of the best seed for each band.

The initial modelling demonstrates interesting results beyond being an starting point for the final fit. This includes chromatic changes in geometry and comparisons between Sy1 and Sy2 cases. Therefore, we present and discuss these results in Sect\,\ref{s:initialBbB}.

\subsection{Band-by-band fit}\label{sec:finalind}

The final best fits for each seed compared to the data can be seen in Fig.\,\ref{fig:observables_final} and the images are shown in Fig.\,\ref{fig:bbb_final}. The final best fitting parameters are plotted in Fig.\,\ref{fig:results_initial} as the Xs.

It can be seen in Figs.\,\ref{fig:bbb_final} and \ref{fig:observables_final} that the different seeds produce visually and interferometrically similar models. The similarity enforces the idea that the seed does not significantly impact the result. It can also be seen in Table\,\ref{tab:results} that the uncertainty for each parameter in each band is far greater than the variations between seed. However, there is some variation between seeds so the model is not completely insensitive to it.

We find that the model provides a good description of the data for each band. The closure phase is reproduced with an error weighted mean residual of $<8^\circ$ in every band. For V$^2$, we have an error weighted mean residual of 1.5$\times10^{-3}$ (7\%) in K-band, 1.5$\times10^{-3}$ (23\%) in L-band, and 4.5$\times10^{-3}$ (16\%) in M-band. However, N-band is the hardest to reproduce, possibly due to the silicate feature, with an error weighted mean residual of 2.9$\times10^{-3}$ (37\%) which could suggest that our description of absorption is incomplete. For example, we do not include the possibility of a foreground component that is reported in \citet{gamez_rosas_thermal_2022}. Alternatively, or in addition, it could be caused by the base temperatures. Because the N-band had a much more tightly defined base temperature for both the disk and the wind, the results suggest that the N-band is far more sensitive to them. This is expected because a 300\,K would peak in the N-band. We chose a flat base temperature with no radial dependence, it is possible that this is an overly simplistic assumption and a radial dependency is required to prevent the N-band becoming too over resolved when the silicate feature obscures the central region. Further modelling would be needed to see if this is due to the dust composition, geometry, temperature laws, or a combination.

\subsection{Polychromatic fit}

The polychromatic fit was initiated from the best starting parameters selected from the band-by-band fit (see Sect.\,\ref{S:M:poly}). We do not enforce edge or face on because only one $inc$ can be found for all bands in a polychromatic fit removing the possibility of the type changing with band. The resulting parameters of the fit can be found in Table\,\ref{tab:results} and Fig.\,\ref{fig:results_initial}. We find that the polychromatic model can well explain the majority of the observable data at every wavelength simultaneously with an edge-on inclination. The V$^2$ (see Fig.\,\ref{fig:polyv2} for best result and Fig.\,\ref{fig:polyv2med} for median result with errors) is well reproduced in K- and L-band with only minor discrepancy in the 100\,m peak which is within 2\,$\sigma$ of the median result. The same is true for the M-band except for a significant discrepancy at the shortest baseline. The shortest baseline V$^2$ in M-band is lower than the next shortest baseline V$^2$ at the same PA, this is not seen at any other covered wavelength. Furthermore, in Extended Data Figure 4 of \citet{gamez_rosas_thermal_2022}, the V$^2$ of every baseline increases or is constant with wavelength except for the shortest baseline. Therefore, we suggest that the shortest baseline has been suppressed from correlated flux loss. This could explain why the shortest baseline has a much higher scatter in the L-band than the other baselines. Finally, the model well reproduces the N-band at 8.7\,$\mu$m and 12.3\,$\mu$m but it is too flat in the silicate feature; i.e. the model is too extended when the central region is heavily obscured. This reinforces the idea that the current obscuration or base temperature description is not sufficient to completely explain the N-band. To check if a different dust composition is responsible, we additionally tried using ISM dust. However, we find that the quality of the fit in N-band is worse when using ISM dust.

\begin{figure*}
    \centering
    \includegraphics[width=\textwidth]{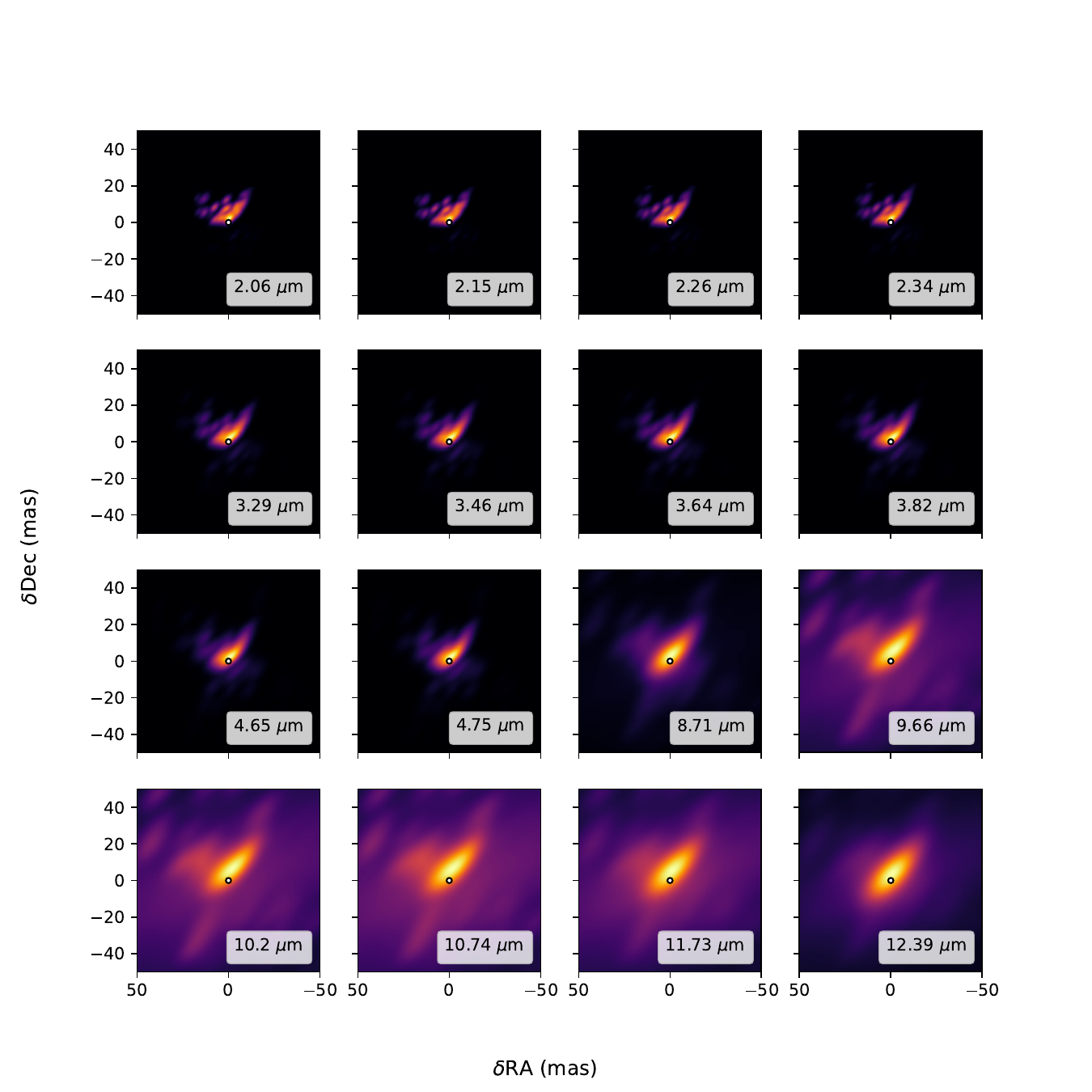}
    \caption{Image of the polychromatic model per wavelength. The white dot is the centre of the model, i.e. the location of the SMBH in an aligned system. All images are colour scaled by a power of 0.6 to highlight faint structure.}
    \label{fig:polyim}
\end{figure*}

\begin{figure*}
    \centering
    \includegraphics[width=\textwidth]{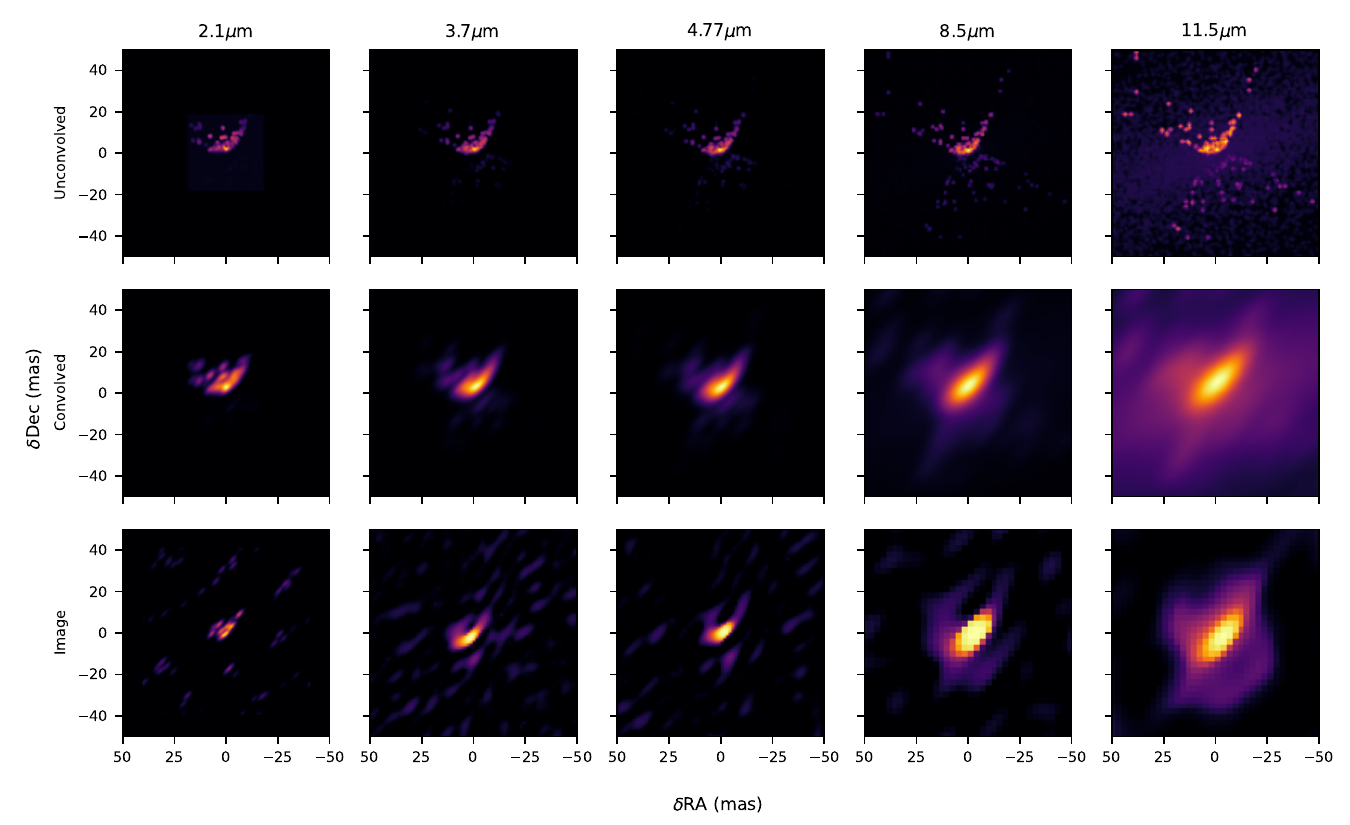}
    \caption{Image of the polychromatic model unconvolved (top row) and convolved with elongated beam (middle row) at the wavelengths of the image reconstructions of \citet{gamez_rosas_thermal_2022,gravity_collaboration_image_2020} and the image reconstructions from these works (bottom row). All images are colour scaled by a power of 0.6 to highlight faint structure.}
    \label{fig:polyimcomp}
\end{figure*}

The overall closure phase (see Fig.\,\ref{fig:polycf} for best result and Fig.\,\ref{fig:polycfmed} for median result with errors) is well reproduced by the best fit model. The closure phase is more sensitive to the small detail in the model, i.e. it is more sensitive the distribution of clumps. Therefore, the errors on the median fit are larger than for the V$^2$. In the K-band, the larger triangles are so sensitive to the absolute position of the clumps that the closure phase is close to uniformly random with different clump distribution for the same geometric parameters. A different seed or a small change in a parameter triggers a different random draw of clump locations which returns a closure phase between $\pm180^\circ$ with near uniform probability for the largest triangle. Hence, the median of these triangles are $\sim0^\circ\pm100^\circ$. However, the triangles with a perimeter of 200\,m or less are well reproduced. We also find that the image in K-band closely resembles the image reconstruction of \citet{gravity_collaboration_image_2020} (see Figs.\,\ref{fig:polyim},\ref{fig:polyimcomp}). We determine that, in this case, the K-band geometry is difficult to interpret alone because of the inherent sensitivity to individual clump locations.



\begin{figure}
    \centering
    \includegraphics[width=0.48\textwidth]{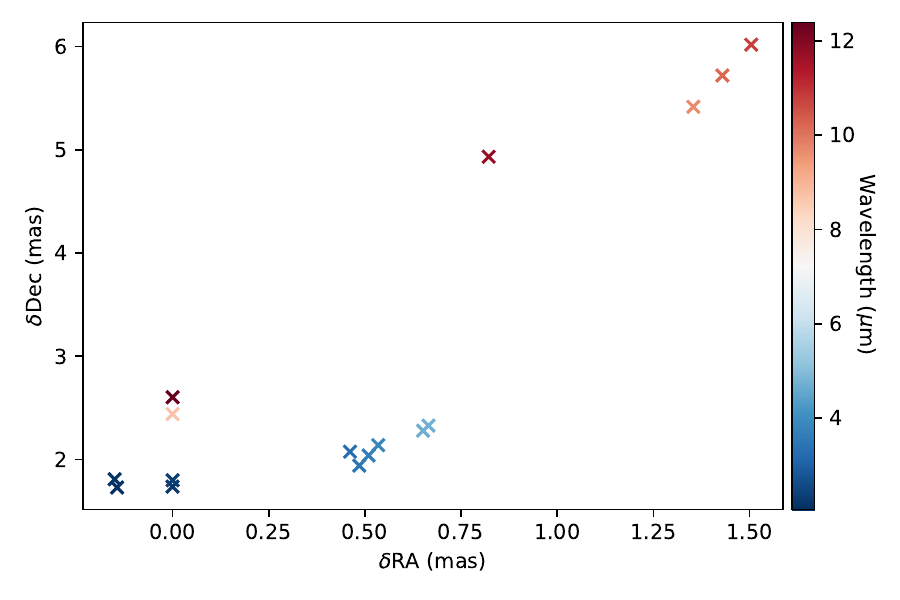}
    \caption{Location of the brightest spot from the polychromatic model with wavelength, (0,0) is the model centre.}
    \label{fig:photocentre}
\end{figure}

Figure\,\ref{fig:polyimcomp} shows the polychromatic fit evaluated at the wavelengths of the image reconstructions of \citet{gamez_rosas_thermal_2022,gravity_collaboration_image_2020}. The model at every wavelength can be seen to be visually similar to the image reconstruction. While the quality of the reproduction of the observables is the indicator of the goodness of the model, the high level of visual coincidence between the model and image reconstructions shows that the model can reproduce the observed brightness distributions at every wavelength simultaneously. No additional distinct, complex, or unique structures are required to explain the visual differences between bands.

One of the questions we aim to tackle is if the assumed alignment of the wavebands in \citet{gamez_rosas_thermal_2022} is accurate in our model scenario. Their aperture photometry, on which their thermal map of the dusty structures was based, was done with pivotal assumptions about the inter-band image alignments. When we compare the model centre of our best fitting model to the brightest pixel over all wavelengths, we find that the photocentre does move (see Fig.\,\ref{fig:photocentre}). Relative to the model centre, we find that the brightest spot moves up to 1\,mas over most wavelengths which is within their aperture. The exception is the silicate feature which moves up to $\sim$5\,mas; however, we do not claim this position with certainty because the fit within the feature is not reliable. Therefore, the assumed alignment between bands in their work is consistent with our model.

\section{Physical interpretation of results and comparisons with literature}\label{sec:physdisc}

{We have presented the modelling results and shown that our model can provide a good reproduction of the observed data. Here we will further discuss the physical implications of the model and make a comparison to different structures observed within the AGN previously.}

\subsection{Molecular gas comparison}

{One of the most direct comparisons which can be made is to the pc scale molecular gas. The dusty structure we observe is mostly gas by mass and such gas can be observed on the pc scales with the Atacama Large Millimeter Array (ALMA). \citet{garcia-burillo_alma_2019} observe CO(3--2), CO(2--1), and HCO$^+$(4--3) in NGC\,1068 with an angular resolution of $\sim50-100$\,mas. This overlaps the VLTI covered scales in the N-band and is well within the spatial extent of our model. {The extent of the line emitting structure, with a major-axis FWHM of $\sim100-250$\,mas depending on the emission line, is also well within our model outer limit ($r_\mathrm{out}\simeq888$\,mas for the polychromatic model) and the within the extent of our best fitting dust disk (FWHM of $\sim$420\,mas) and can therefore be compared.}} They find that each molecular gas emission line is elongated with an average PA of 113$^\circ$ and, assuming the gas is in a disk, find an upper limit on the inclination of $30^\circ-40^\circ$. The result of the molecular emission line analysis is in complete agreement with the disk component of the polychromatic model with an inclination of $16^\circ\pm1^\circ$ and a PA of $112^3_2{}^\circ$ (PA of $22^3_2{}^\circ$ + $90^\circ$). These gas lines predominantly tracing the disk component and not the wind is consistent with expectations \citep[see][]{honig_redefining_2019} as the gas lines trace denser regions and the molecular gas is mostly destroyed by radiation in the wind.

{\citet{gallimore_high-velocity_2016} observe CO(6--5) {with ALMA and also find a disk at $112^\circ$ consistent with the other ALMA observations and our work.} Furthermore, they do succeed in observing this gas line in an outflowing structure. This structure is on mas spatial scales with a PA of $33^\circ$. This is slightly offset by $11^\circ$ with our model but it is in the same direction. An $11^\circ$ offset is also on the same order as the combined uncertainty of our model and the estimated uncertainties in the CO(6--5) outflow direction from the beam size.}

\subsection{ALMA observed dust}

{ALMA has also observed dust continuum in NGC\,1068 \citep{garcia-burillo_alma_2016}. The observation had an angular resolution of $70\times50$\,mas at 432\,$\mu$m meaning the spatial scale overlaps with our N-band. The dominant dust temperature probed by ALMA should be cooler and, therefore, disk dominated at these scales. They find the dust has a PA of $142^\circ\pm23^\circ$ which by itself is consistent with our disk. Furthermore, direct comparison between our model at longer wavelengths and their Fig.\,1\,b and c contain additional possible aligning features. The southwest extension on larger scales in their Fig.\,1\,b is in the same direction as our wind. Their Fig.\,1\,c, which shows a slightly higher resolution image, shows an extension much closer to our model disk with a Northern spur that lies along one edge of the outflowing wind. We conclude that our result is completely consistent with the cooler ALMA observed dust.}

\subsection{Radio and H$_2$O maser emission}

{
Both \citet{gravity_collaboration_image_2020} and \citet{gamez_rosas_thermal_2022} make comparisons to the H$_2$O maser emission. If we assume that the 5\,GHz position of S1 in \citet{gallimore_high-sensitivity_2023} aligns with the centre of our polychromatic model (i.e. the "central engine" position), we find that the maser disk (their R4+G1 maser group) traces the rim of the obscuring disk and the other groups, which they attribute to spiral arms, trace the outer wall of the dusty outflow. The blueshifted groups coincide with the other side of the outflow to the redshifted. We plot the maser emission against the unconvolved polychromatic model at $2\,\mu$m in Fig.\,\ref{fig:masers}. This is an intriguing coincidence but if the groups coincidental with the wind are associated, then we might expect the outflowing cone inclined toward the observer to coincide with the blueshifted emission whereas it is instead redshifted. We also find that with this alignment, the 22\,GHz continuum emission that matches the brightness distribution in a similar manner to \citet{gamez_rosas_thermal_2022}. Aligning by the dynamic centre of the maser disk instead of the 5\,GHz peak produces similar alignments. Compared to \citet{gravity_collaboration_image_2020}, \citet{gamez_rosas_thermal_2022}, and \citet{gallimore_high-sensitivity_2023}, our alignment is close to model 4 of  \citet{gravity_collaboration_image_2020} and that of \citet{gamez_rosas_thermal_2022}. It also agrees with the scenario presented in Sect.\,5 of \citet{gallimore_high-sensitivity_2023} where the maser disk (R4+G1) aligns with the obscuring disk.}

\subsection{Narrow line region}

{In the unified scheme of AGN, the Narrow Line Region (NLR) outflows along the polar direction, similar to the model dusty wind. Whether the NLR is coupled to the dust outflow or not, in the simplest case of an aligned system they should both outflow in the same direction. However, it should be noted that the NLR is currently only studied on much larger scales than we probe in this work making comparisons less precise.}

{Using a biconical outflow model, \citet{das_kinematics_2006} finds a NLR PA of $30^\circ$, inclination of $5^\circ$, and an outer opening angle of the wind of $80^\circ$. However, they probe $\sim4$" angular scales which is much larger than our FoV. \citet{poncelet_dynamics_2008} finds that on smaller $\sim1$" scales that the NLR is consistent with a bicone of PA $=10^\circ$, an inclination towards the observer of $\sim10^\circ$, and an opening angle of $\sim82^\circ$ for the cone. Given the differences between the works, the difference between the spatial scales probed, and the inherent complexity of NGC\,1068 we conclude that our wind structure, with a PA $=22^3_2{}^\circ$; an inclination of $16^\circ\pm1^\circ$; and an opening angle of $90^\circ\pm8^\circ$, is consistent with the NLR.}

\subsection{Extended hot dust and scattered light}

{In Sect.\,\ref{sec:over-resolved}, we discussed the extended $\sim100$\,mas structure observed by \citet{weigelt_diffraction-limited_2004} in relation to the hot over-resolved component seen in our model and by \citet{gravity_collaboration_image_2020}. However, \citet{weigelt_diffraction-limited_2004} produced K- and H-band image reconstructions with sub-hundred mas resolution. Our main K-band emitting structure would be unresolved in their images; however, should the hot over-resolved component indeed be scattered light then the scattering medium would be the outflowing dust and the geometry should be related to our wind structure. The extended structure of their images, particularly in H-band, show striking resemblance to our wind structure. Each branch of the X-shape in their Fig.\,3 H-band image are closely aligned with the edges of our wind. Their K-band image shows a narrower, less X-like, elongation which aligns with our wind. Additionally, the wind and disk well align with the polarised light H- and K-band structures seen in \citet{gratadour_polarimetric_2015}. This coincidental alignment could indicate that the extended K- and H-band structure, and therefore our over-resolved hot component, is indeed from scattered accretion disk light. However, a more in-depth study of the hundred mas scales is needed to make firm conclusions about the origin of this emission.}

\subsection{Comparison summary}

{From these comparisons and our model, we can place the VLTI observed structure in the bigger picture. We conclude that the data of NGC\,1068 is consistent with the AGN containing a dusty obscuring molecular gas disk that obscures the dust sublimation region and the hotter part of the disk. From this disk dust is ejected into a wind coincidental with the outflowing NLR and dominates the K-, L-, M-, and N-band emission in the central pcs but between $50\%-70\%$ of the K-band flux on hundred mas scales is not from AGN heated thermally emitting dust. The wind dust may then act as a scattering surface which scatters the accretion disk emission contributing to at least part of the missing K-band emission. Finally, the less obscured cooler dust further out in the disk then starts to become more significant in emission in the sub-mm regime seen by ALMA.}

\section{Comparison of models, improvements, and future prospects}\label{s:Discussion}

We have presented the results of the polychromatic modelling of NGC\,1068. There are some notable features which are worth further consideration which we will discuss here.

\subsection{Comparisons between models}

Initially we expected that the band-by-band modelling would be a better reproduction of the data than the polychromatic modelling because it would allow more complex chromatic changes between bands than our model can describe to be approximated by changes in geometry. When we compare the band-by-band results to the polychromatic results, we find that they reproduce the data comparably per wavelength. In the N-band, the polychromatic model actually provides a better visibility reproduction and a more visually similar brightness distribution to the image reconstructions (see Fig.\,\ref{fig:polyimcomp}) which we attribute to the wider wavelength range providing better constraints on the inner geometry. Because the one polychromatic model uses fewer free parameters than three band-by-band models to explain the same data with a similar quality, we can conclude that the polychromatic model is a better general description. This is an encouraging sign that our model is not missing any significant components. Perhaps more importantly, it validates the registration of emission in each band to common components.

When we compare the recovered parameters between the two modelling methods we find good agreement in most cases. We use the band-by-band as prior information for the polychromatic fit and, therefore, we would expect the results to remain similar unless a wider wavelength range provides information not available within a single band. The only parameters we find contain a significant discrepancy are $\gamma_\mathrm{off}$ and $f_w$, the break radius of the disk temperature power law and the fraction of dust clumps in the wind component, respectively. We adjusted $\gamma_\mathrm{off}$ in the final model to determine the effect on the observables. We find that the parameter is poorly constrained in the edge-on case and the offset is in reality insignificant. We suggest that this is because the disk is mostly self obscured and, therefore, the inner break location is not seen. We find that the discrepancy in $f_w$ is significant. Using $N_p$ to convert $f_w$ into an absolute number of points in the wind, we still find a discrepancy. This means the polychromatic model contains less dust in the wind. We attribute this difference to the additional over-resolved component. The polychromatic model prefers a more central distribution of points in the wind (higher $\alpha_w$) which results in fewer points outside the image and fewer points needed to construct similar central regions. The fewer points outside the image leads to less over-resolved flux in the model which is compensated by the additional over-resolved parameter. This is additionally evidenced by the fact that $V_0$ does not increase despite the addition of $BB_\mathrm{hot}$. In the band-by-band fits, the achromatic over-resolved component $V_0$ was compensating for both the achromatic and chromatic over-resolved structure.

The necessity of $V_0$ in the polychromatic fit is also interesting. This suggests that there is an over-resolved structure responsible for $35\pm7$\% of the flux at every wavelength. However, we see in the band-by-band fit that $V_0$ seemingly peaks in L-band. It is possible that instead of an achromatic component, we have cool dust component of a few hundred\,K causing more over-resolved flux in M- and N-band than L-band {and that the hot over-resolved component is brighter but is being partially compensated by $V_0$}. However, this is difficult to conclude at this time. In future work, the addition of the total flux when modelling would provide an additional constraint to such a component. Due to the interactions between galactic material, the jets, and the winds, such a cool component is conceivable \citep[e.g. AGN heated 475\,K dust at 450\,mas from the centre seen with NACO:][]{rouan_hot_2004}.

\subsection{Silicate absorption feature}

The polychromatic modelling with the disk+wind scenario produced an answer that can reproduce the V$^2$ and $\Phi_c$ with a error weighted mean residual of 0.0035 (20\%) and 11$^\circ$, respectively. The largest deviation between the model and the data is within the silicate feature near 9.7\,$\mu$m. Future iterations of the model should allow for different dust compositions to test if this is caused by composition. In this work, we used the olivine and amorphous carbon dust species used in \citet{gamez_rosas_thermal_2022}. In future work, using the \textsc{dustem} python package \citep{compiegne_global_2011} would allow for greater flexibility in dust absorption mixes.

\subsection{Future prospects}

In this work we used our model to describe one AGN; however, we developed the model with the intent to keep it general rather than specific to NGC\,1068. Primarily, this was to test if NGC\,1068 could be explained by a simple generic AGN geometry but it also allows the model to have other applications. Here we will give some prospects towards what could be done in the future.

\subsubsection{The over-resolved components}

{In future work we will investigate the physical nature of the over-resolved components. This can be done in part by including the SED and correlated fluxes which will provide additional constraints on the absolute flux of the components. The SED is particularly necessary to distinguish if the hot-component is indeed hot or instead non-thermal and if it is thermal what temperature it is. Furthermore, high angular resolution imaging in the IR will be studied which may resolved these structures. They could also potentially be resolved through the shorter baselines provided by VLTI AT observations, which we did not include in this work due to complex mismatched FoV effects, or through in-progress studies with shorter baseline LBTI \citep[Large Binocular Telescope Interferometer][]{hinz_large_2002} observations. Additionally, approved aperture masked data from 8\,m class telescopes in the IR will provide access to the scales of the over-resolved structures (priv. comm.).}

{These components are interesting to investigate for NGC\,1068. However, it is unclear whether they are required to explain VLTI observed AGN in general. AGN are extended objects in the IR and often have over-resolved structure but if the IR is dominantly centrally heated dust emission then the over-resolved structure would inherently be included in the model. If the additional components needed for NGC\,1068 are a common feature related to the AGN dust (e.g. by scattering), it could be included in the model as a physical feature of the dust clumps themselves rather than a scaling factor. Our model will need to be applied to other AGN to determine if this is the case. Furthermore, this structure is over-resolved in the relatively local NGC\,1068 but in more distant AGN such structures may become resolved.}

\subsubsection{Radiative transfer simulations}

A line of investigation that we do not fully cover in this work is the physicality of the model. Our model is a "toy" model in the sense that we base our geometry and assumptions on previous work and physics but we do not perform detailed physical simulations such as hydrodynamics or full RT; i.e. we calculate dust emission and absorption in a physical manner but we set dust temperatures, densities, and distribution parametrically. In a future work, we will collect our toy model parameters and translate them into an RT simulation. The goal of such a work would be to see if our model geometry and dust densities could reproduce our derived distribution of temperatures and brightness in RT. If so, we may better constrain the dust properties in a more "physically realistic" manner. Such a work would allow us to place better constraints on the realism of the properties we derive and test if our model could be a less computationally expensive way to obtain a geometry and dust composition which can be followed up with a more expensive RT simulation once constrained.

\subsubsection{A larger AGN sample}

{In future work we will utilise the model on other MATISSE-GRAVITY observations of Sy1 and Sy2 AGN with less coverage than NGC\,1068. Many AGN have been observed with MATISSE and GRAVITY that do not have sufficient \textit{uv} coverage to perform image reconstructions. Our model has proven sufficient to explain the observables of the NGC\,1068 "test case" and it produces models visually similar to the image reconstructions. Therefore, we can now apply the model to AGN for which image reconstruction is not possible.}

{NGC\,1068 is very well observed and we utilise that through a simple but still relatively complex model when comparing the number of free parameters to models used on less well covered AGN \citep[e.g.][]{leftley_resolving_2021, gravity_collaboration_resolved_2020, lopez-gonzaga_mid-infrared_2016, burtscher_diversity_2013}. 25 free non-error parameters, equivalent to a fixed point source and four free elongated 2D Gaussians, is not excessive for NGC\,1068 as demonstrated in \citet{gamez_rosas_thermal_2022}. For AGN observed with only one or two snapshots with MATISSE or GRAVITY alone, some parameters may need to be fixed to prevent over fitting and degeneracy. However, a lower required \textit{uv} coverage is one advantage of modelling over image reconstruction and an additional advantage of a chromatic model is the significant increase in observational data in a fit. This leads to better constraints for the same \textit{uv} coverage as demonstrated by the improvement between the band-by-band and polychromatic fits. This should allow all parameters to be free for smaller numbers of MATISSE snapshots than achromatic models. For a large survey, the complexity constraints from the data coverage will need to be investigated per object.}

\subsubsection{Improvements}

While RT is a next step in modelling of NGC\,1068, our model can be used on other objects. For this, there are improvements that can be made in the toy model itself. For example, it is clear that the model could benefit from the inclusion of more flexible dust compositions. It has been designed such that different dust species can be included easily by providing their optical cross sections per gram. As such we can incorporate existing dust libraries, e.g. \textsc{dustem} \citep{compiegne_global_2011}, to give users greater flexibility.

Furthermore, because of dust is described by probability distributions, new geometries can be included as needed and added arbitrarily. We will streamline this process for ease of use with the goal of including the model as an option within \textsc{oimodeler}\footnote{\url{https://oimodeler.readthedocs.io/}}.

\subsubsection{Generalisation outside AGN}

We focus on AGN in this work but our model is one of dusty structures. Many dusty objects in astronomy can be described by a disk or a disk+wind. As such, the model can be applied to objects such as some B[e] stars, young stellar objects, and proto-planetary disks. The list can be further extended once we allow user defined geometry.

\section{Conclusions}\label{s:Conclusions}

We present our successful attempt to reproduce the IR interferometric observables of NGC\,1068 from GRAVITY and MATISSE simultaneously with a polychromatic model capable of covering the full K-band to N-band range. This represents a breakthrough in very broad band modelling of dusty structures at very high angular resolution.

We find that a simple clumpy disk and hyperbolic wind composed of thermally emitting non-ISM dust can reproduce the squared visibility (V$^2$) and closure phase ($\Phi_c$) with an error weighted mean residual of 0.0035 (20\%) and 11$^\circ$, respectively, with only one additional ad hoc structure. {The resulting PA, inclination, and wind opening angle are consistent with those derived from observations of molecular gas \citep{garcia-burillo_alma_2019}, ALMA observed dust \citep{garcia-burillo_alma_2016}, polarised IR light \citep{gratadour_polarimetric_2015}, IR speckle interferometry \citep{weigelt_diffraction-limited_2004}, radio and maser emission \citep{gallimore_high-sensitivity_2023}, and the NLR \citep{poncelet_dynamics_2008, das_kinematics_2006} which strongly supports that the IR emission can be explained by the disk+wind iteration of the unification scheme.} 
The additional structure is an over-resolved hot or {non-thermal component ($>>2000\,$K, approximated as a 20000\,K black body), e.g. a central sub-arcsecond star forming region or scattered light from the central accretion disk.} Furthermore, our obscuration, from both dust clumps and a smooth dust disk, reproduces the data well but over-obscures within the N-band silicate feature; in future work we should investigate if a change in geometry or composition could improve this, e.g. using a puffed up inner rim in the disk or changing the dust composition. 

When we model K- and L-band individually, we find it difficult to distinguish between two cases for the recovered system: an inclined ring (Sy1 like) or an edge-on wind with an obscured central region (Sy2 like). When fitting all wavelengths simultaneously, we find the model strongly favours a Sy2 - like dust distribution where the central region is obscured, this supports the conclusions of \citet{gamez_rosas_thermal_2022} and demonstrates the additional constraining power of chromatic information when modelling IR interferometric observations. The model also yields an alignment between K-, L-, M-, and N-bands, allowing us to link the structures in the image reconstructions of \citet{gravity_collaboration_image_2020} and \citet{gamez_rosas_thermal_2022}. Our model interprets the bright regions of each band as the wind whilst the sublimation region remains fully obscured at all observed wavelengths, this supports the unified theory of AGN presented by \citet{antonucci_unified_1993} and the disk+wind interpretation presented in \citet{honig_redefining_2019}. We conclude that the disk+wind unified model of AGN is a good description of the inner structure of NGC\,1068 in the IR which can be translated into a full radiative transfer model to further investigate the dust properties.

In this work we have demonstrated simple chromatic modelling as a powerful tool for IR interferometric data. In future work we will extend our geometry for NGC\,1068 to a full RT simulation and apply our versatile modelling approach to a larger sample of AGN. This will further test our approach and construct a sample of AGN for which we have a good description of their dust structure. However, our technique is not limited to AGN. In the future, we aim to provide our model as a general tool for interferometric studies of dusty structures (e.g. B[e] stars, young stellar objects, and proto-planetary disks).

\begin{acknowledgements}
We would like to thank the referee for the kind comments and suggestions which helped improve this work.

This work was supported by French government through the National Research Agency (ANR) with funding grant ANR AGN\_MELBa (ANR-21-CE31-0011) and the Universit\,e C\^ote d'Azur JEDI investment in the Future project managed by the  under the reference number ANR-15-IDEX-01.

This research has made extensive use of NASA’s Astrophysics Data System; the SIMBAD database and VizieR catalogue access tool, operated at CDS, Strasbourg, France; and the python packages \textsc{astropy}, \textsc{emcee}, \textsc{scipy}, \textsc{galario}, and \textsc{matplotlib}.
\end{acknowledgements}



\bibliographystyle{aa}
\bibliography{referencesla} 




\begin{appendix}

\section{Band-by-band results - Additional}

\subsection{Initial band-by-band fit}\label{s:initialBbB}

Here we present the initial results of the band-by-band modelling for the Sy2 case and we present the Sy1 case in Appendix\,\ref{sec:faceind}. The initial fits contain some interesting and useful results beyond providing a starting point for the second run and, therefore, we present them here. It should be noted that we present these results because their trends and scatter inform us about structure we might otherwise overlook, the dependency of the model on the seed, and how strongly defined is each initial parameter for the final fit. These results should not be used for further analysis, instead the results in Sect.\,\ref{sec:finalind} should be used. The initial run is performed to determine a reliable starting position for the final run as a method to fully sample the parameter space in a more time efficient manner. It is not run for a sufficient length of time to determine reliable errors from a uniform start for our case.


\subsubsection{Sy2 case}\label{sec:endind}

\begin{figure*}
    \centering
    \includegraphics[width=\textwidth]{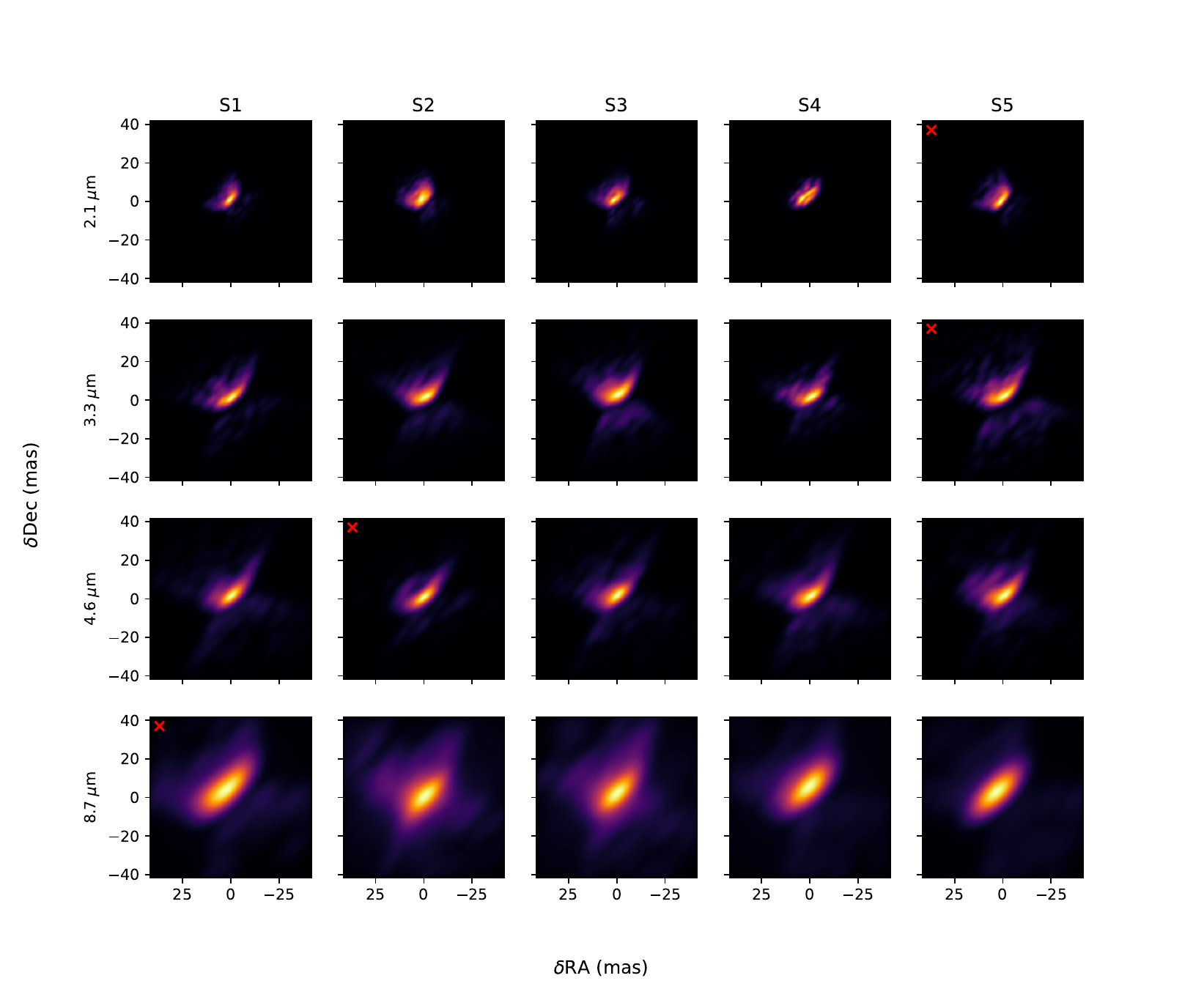}
    \caption{Best model for each band (row) and each seed (column) for the initial Sy2 band-by-band fit as determined by the maximum likelihood. The best seed for each band is marked with a red x. The K-, L-, M-, N-bands are evaluated at 2.1\,$\mu$m,3.5\,$\mu$m,4.7\,$\mu$m, and 8.5\,$\mu$m, respectively. All images are normalised and given a 0.6 power colour scaling to match that of \citet{gamez_rosas_thermal_2022} for easy comparison.}
    \label{fig:bbb_initial}
\end{figure*}

The initial run, without a well defined starting position, produced results that were very consistent between seeds in the L-, M-, and N-bands suggesting that the underlying geometry can be found in these bands for any seed. 
The K-band produced a wider array of results which suggests that it could be more sensitive to the individual clump positions, and therefore the random seed, than the longer wavelengths. The best model from the initial fit for each seed and band can be seen in Fig.\,\ref{fig:bbb_initial}. The parameters for the initial fits for each seed can be found in Fig.\,\ref{fig:results_initial}. We compare the data from each best fit model to the data in Fig.\,\ref{fig:observables_initial}.

\begin{figure}
    \centering
    \includegraphics[width=0.48\textwidth]{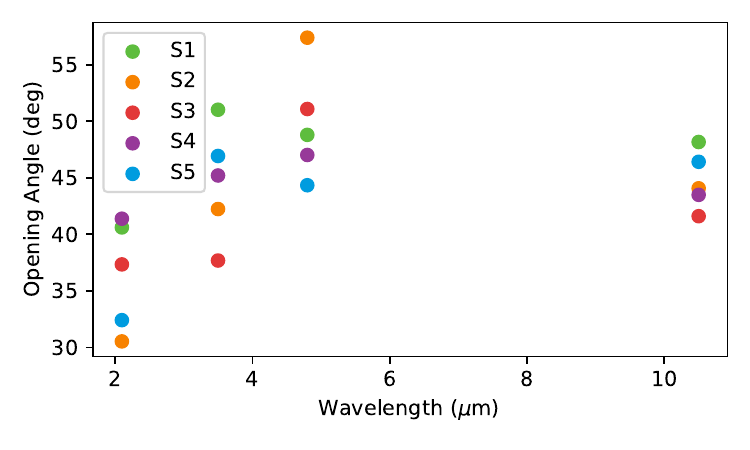}
    \caption{Opening angle of the initial band-by-band fit in the Sy2 case (Sect.\,\ref{sec:endind}) against wavelength.}
    \label{fig:opang}
\end{figure}

Most parameters were consistent between bands within the scatter of the different seeds; however, there are some exceptions. Of particular interest are $a_w$, $b$, $ang$ and $inc$. $inc$, the inclination angle, remains consistent within the inter-seed scatter between the all bands at an average value of $\sim15^\circ$ which is close to the NLR geometry inferred inclination of $10^\circ$ \citet{poncelet_dynamics_2008}. 
$ang$, the PA of the cone or the system/polar axis, is consistent within inter-seed scatter in the L-, M-, and N-bands at $24^\circ\pm 5^\circ$ which agrees with the system axis from \citet{asmus_subarcsecond_2016} of $28^\circ\pm20^\circ$ measured from the NLR, radio, masers, and mid-IR imaging. $ang$ in K-band is highly scattered but consistently higher. Although $ang$ is close between L-, M-, and N-band, there is a tentative positive relation with wavelength that could be indicative of a chromatic anisotropy in the brightness distribution of the wind. For example, anisotropic illumination of the wind by the accretion disk causes one side to be hotter than the other which would be more apparent at shorter wavelengths and smaller spatial scales. Our model does not allow for anisotropic illumination and, therefore, would manifest as a rotation. 
$b$ tentatively increases with wavelength whereas $a_w$ decreases more strongly. $b$ and $a_w$ are both parameters that define the hyperbolic wind shape and a hyperbolic line of $\frac{z^2}{a_w^2}-\frac{x^2}{b^2}=1$ approaches $z=\frac{a_w}{b}x$ at large $x$. If we define the opening angle of the wind as the angle of the approached line from the polar axis, we can define it as $\arctan\left( b/a_w\right)$. Then the wavelength dependence of $b$ and $a_w$ instead suggest that the opening angle of the wind increases with wavelength 
from an opening angle in the K-band of $\sim$40$^\circ$ 
to $\sim$50$^\circ$ 
in the L-band. 
We show this in Fig.\,\ref{fig:opang}.

Finally, we find that $V_0$ is consistently lower in K-band which is expected from our initial assumption we would require an over-resolved hot component that is not included in the band-by-band fits. It is consistent with the over-resolved fraction of 0.5 found in \citet{gravity_collaboration_image_2020} when performing the image reconstruction in K-band.



The presence of the silicate feature makes the N-band observables difficult to reproduce without a wider wavelength coverage. We suspect that this is because much of the N-band is heavily obscured and therefore much of the central geometry is not available; conversely, the wider FoV makes more of the extended geometry available. This results in the different seeds providing similar $ang$ and geometry for the winds but the disk temperature distribution is unconstrained. 

Compared to the other bands, $n_d$ is high and better constrained in the N-band. This is expected because of the presence of the silicate feature which is a strong constraint on the obscuration. However, $n_c$ is no better constrained or higher in the N-band than the other bands. This suggests that the smooth obscuration is relatively more impactful at longer wavelengths, possibly due to the lower resolution causing reduced distinction between smooth and clumpy structure. When translating the recovered $n_d$ and $n_c$ into $\tau$ along an equatorial LOS towards the model centre, we find $n_c$ is the dominant contribution to the disk obscuration due to the number of clumps in the LOS. However, $n_d$ is still significant for some LOS. Particularly if the LOS is through a region with fewer clumps.

The initial L-, M-, and N-band images produced by the models (see Fig.\,\ref{fig:bbb_initial}, the N-band is evaluated at 8.5\,$\mu$m) are all quite consistent between seeds and more importantly with the image reconstructions of \citet{gamez_rosas_thermal_2022}. The resemblance between the two methods already confirms that the brightness distributions found in \citet{gamez_rosas_thermal_2022} can be reproduced by a disk+wind geometry.





At first glance, the K-band is seemingly less comparable to the image reconstruction of \citet{gravity_collaboration_image_2020}; although, the image produce by seed S4 is visually similar with a ring-like structure and surrounding clumps. 
We find that the K-band V$^2$ are dominated by a North-East to South-West binary-like structure with closure phase information introduced by a significantly clumpy structure. This is what is seen in the image of \citet{gravity_collaboration_image_2020}, there is a bright complex North-East component, the ring-like structure, and a dim Southern clump. We suggest that because the true brightness distribution is seemingly clumpy and that for a clumpy structure, due to finer angular resolution, the relative positions of the clumps would more important in K-band than in longer wavelengths, it is unlikely that our model, particularly in the initial fit, would find a perfect solution for a particular seed. This is supported by the higher variety of images and parameters in the K-band indicating a higher sensitivity to the seed. Therefore, the more likely solution for the model to discover is dominated by the easy to reproduce binary-like structure which is a good description of the V$^2$ and the $\Phi_c$ at smaller triangle perimeters (<220m). As such, we often find a bright dense cluster of clumps and a faint cluster, a solution that can be easily reproduced with any seed. This bright-dim spot structure can also be reproduced with a ring similar to the one in \citet{vermot_inner_2021}. 




\subsubsection{Sy1 case}\label{sec:faceind}

The initial run for the Sy1 case starts from the same parameters as the Sy2 case except that $\phi$ must be less than $inc$. We note that this initial run will not be likely to reproduce case of \citet{vermot_inner_2021} because it would require an $ang$ of 180$^\circ$ offset from the initial position. This Sy1 start is specifically to test the model 3 and model 4 of \citet{gravity_collaboration_image_2020}. The best fits for each seed compared to the data can be seen in Fig.\,\ref{fig:observables_initial_face} and the images are shown in Fig.\,\ref{fig:bbb_initial_face}.

The Sy1 case, with the initial parameters given, converged to Sy2 like results. The model preferred inclinations that were as close to obscured as allowed or positioned such that the wind acted as the obscurer for the central region. The best fits produced were the same order of magnitude as the Sy2 case when comparing likelihood; however, they were consistently of lower likelihood.

\subsubsection{Rotated Sy1 case}\label{sec:ring}

To test the ring model of \citet{vermot_inner_2021}, we perform the same initial modelling as the Sy1 case but offset the start of $ang$ by 180$^\circ$. It is important to note that  \citet{vermot_inner_2021} explains the ring as a distinct structure, such as a tidally disrupted clump, separate from the unification scheme. Our model can produce a ring that produces a similar brightness distribution but this ring will be the sublimation region, not a distinct structure.

For this case, we recover the ring structure from \citet{vermot_inner_2021} in K-band and in L-band. In M-Band a ring is still recovered but closer to edge-on and in N-band, we recover the Sy2 case. We conclude that the ring fit is a valid solution in K- and L-band individually (see Fig.\,\ref{fig:ringmodel}). However, when we perform a polychromatic fit of all bands begun from the L band ring fit we find that it is not valid for when simultaneously explaining all wavelengths with our geometry. The polychromatic ring fit does not reproduce the observables as well as the Sy2 model, particularly in the N-band, as can be seen in Figs.\,\ref{fig:ringmodelv2} and \ref{fig:ringmodelphi}.

\section{Technical checks}

\subsection{Dust mass}

In our model, we calculate the dust obscuration from clumps and smooth structure. This means we can translate this into a dust density and mass. While this will be inaccurate without the SED, we can check that we do not use unphysical amounts of dust for our best fitting model. In the band-by-band fits, $n_c$ varies between $3-4$; however, within a band the value is significantly more constrained. This is most likely due to the dust composition being poorly constrained by a single band and therefore $n_c$ becomes semi-degenerate with $p_g$. In Fig.\,\ref{fig:dustcomp} we show the cross section per gram of the two materials in our dust mix, it can be seen that outside the silicate absorption feature the two materials have differences of up to two orders of magnitude. Therefore, for example, in K-band you can achieve the same absorption with pure amorphous carbon and pure olivine dust if you have $\sim45$ times more dust by mass, equivalent to a change in $n_c$ of 1.65. Ergo, for better constraints on a clumps mass we use the result of $n_c=-3.9^{0.2}_{0.3}$\,$\log_{10}$(g\,cm$cm^{-2}$\,mas$^{-1}$) and $r_c=1.1^{0.1}_{0.1}$\,mas from the polychromatic model. Using a distance to NGC\,1068 of 14.4\,Mpc from the NASA/IPAC Extragalactic Database\footnote{The NASA/IPAC Extragalactic Database (NED) is funded by the National Aeronautics and Space Administration and operated by the California Institute of Technology.} (NED), we can use a conversion from mas to cm of 2.15$\times10^{17}$\,cm\,mas$^{-1}$. This leads to the volume of a clump being $42\times10^{51}$\,cm$^3$ which gives a mass per clump of $0.016^{0.012}_{0.008}$\,M$_\odot$. This is equivalent to a total dust mass in the disk of $270^{210}_{140}$\,M$_\odot$, including the uncertainties from $N_p$ and $f_w$. \citet{garcia-burillo_alma_2019} measure a gas mass of the molecular torus of $\sim1.5\times10^5$\,M$_\odot$ which would make the disk $\sim$0.2\% dust. 
For comparison, this is below ISM values and, therefore, our model is not requiring an unphysical amount of dust.

\section{Figures and Tables}

\begin{landscape}
\begin{table}
\caption{Model Results}
\normalsize
    \centering
    \begin{tabular*}{\linewidth}{@{\extracolsep{\fill}} ccccccccccccccccc}\hline
        Band&Seed&$\alpha$&$\alpha_\mathrm{off}$&$\beta$&$\phi$&$b$&$r_\mathrm{in}$&$\alpha_r$&$\gamma_r$&$a_w$&$n_d$&$\gamma$&$\gamma_\mathrm{off}$&$inc$&$ang$&\\[1.5pt]\hline
        &&&&mas&deg&&mas&$r_\mathrm{in}$&$r_\mathrm{in}$&&$log_{10}$($\frac{\mathrm{g}}{\mathrm{ cm}^2}$)&&&deg&deg&\\[1.5pt]\hline\hline
        All&SP1&$0.32^{0.03}_{0.03}$&$0.51^{0.05}_{0.04}$&$180.0^{20.0}_{20.0}$&$29.0^{3.0}_{2.0}$&$1.6^{0.2}_{0.1}$&$3.0^{0.3}_{0.3}$&$4.1^{0.3}_{0.3}$&$3.5^{0.3}_{0.4}$&$1.6^{0.1}_{0.1}$&$-2.3^{0.1}_{0.1}$&$2.2^{0.3}_{0.2}$&$0.14^{0.01}_{0.01}$&$16.0^{1.0}_{1.0}$&$22.0^{3.0}_{2.0}$\\[1.5pt]\hline

        K&All&$0.43^{0.035}_{0.031}$&$0.95^{0.07}_{0.083}$&$97.0^{8.1}_{7.7}$&$22.0^{1.8}_{1.6}$&$1.2^{0.083}_{0.094}$&$3.3^{0.2}_{0.23}$&$4.2^{0.31}_{0.32}$&$3.0^{0.25}_{0.27}$&$1.9^{0.13}_{0.14}$&$-3.1^{0.16}_{0.14}$&$1.5^{0.12}_{0.11}$&$0.99^{0.099}_{0.092}$&$17.0^{1.1}_{1.2}$&$52.0^{2.9}_{2.9}$\\[1.5pt]
        &S1&$0.43^{0.026}_{0.029}$&$0.95^{0.047}_{0.059}$&$98.0^{5.5}_{7.3}$&$22.0^{1.6}_{1.3}$&$1.2^{0.05}_{0.053}$&$3.3^{0.22}_{0.22}$&$4.2^{0.22}_{0.21}$&$3.0^{0.18}_{0.23}$&$1.9^{0.12}_{0.097}$&$-2.9^{0.1}_{0.1}$&$1.5^{0.11}_{0.1}$&$1.0^{0.074}_{0.056}$&$16.0^{0.75}_{0.79}$&$51.0^{2.5}_{2.5}$\\[1.5pt]
        &S2&$0.43^{0.02}_{0.019}$&$0.95^{0.073}_{0.068}$&$98.0^{5.9}_{7.1}$&$22.0^{1.6}_{1.3}$&$1.2^{0.088}_{0.089}$&$3.3^{0.13}_{0.2}$&$4.1^{0.23}_{0.19}$&$3.0^{0.22}_{0.19}$&$1.9^{0.12}_{0.13}$&$-2.9^{0.12}_{0.13}$&$1.5^{0.081}_{0.068}$&$0.99^{0.084}_{0.11}$&$17.0^{0.96}_{0.79}$&$51.0^{2.2}_{2.3}$\\[1.5pt]
        &S3&$0.44^{0.046}_{0.034}$&$0.92^{0.076}_{0.091}$&$96.0^{9.6}_{8.2}$&$21.0^{1.5}_{1.8}$&$1.1^{0.1}_{0.083}$&$3.3^{0.25}_{0.26}$&$4.3^{0.38}_{0.33}$&$3.0^{0.28}_{0.35}$&$1.9^{0.16}_{0.16}$&$-2.8^{0.17}_{0.17}$&$1.5^{0.13}_{0.16}$&$0.99^{0.11}_{0.093}$&$17.0^{1.2}_{1.3}$&$52.0^{2.6}_{2.9}$\\[1.5pt]
        &S4&$0.44^{0.041}_{0.046}$&$0.96^{0.083}_{0.086}$&$96.0^{7.8}_{9.2}$&$21.0^{2.3}_{1.9}$&$1.2^{0.093}_{0.099}$&$3.3^{0.19}_{0.2}$&$4.2^{0.38}_{0.48}$&$2.9^{0.35}_{0.2}$&$1.9^{0.15}_{0.17}$&$-2.9^{0.2}_{0.15}$&$1.6^{0.16}_{0.16}$&$0.97^{0.13}_{0.097}$&$16.0^{1.5}_{1.3}$&$52.0^{3.5}_{3.5}$\\[1.5pt]
        &S5&$0.44^{0.038}_{0.035}$&$0.96^{0.083}_{0.11}$&$100.0^{9.6}_{9.2}$&$22.0^{2.3}_{1.6}$&$1.2^{0.1}_{0.082}$&$3.3^{0.22}_{0.24}$&$4.1^{0.34}_{0.31}$&$3.0^{0.22}_{0.3}$&$1.9^{0.12}_{0.15}$&$-2.9^{0.15}_{0.16}$&$1.6^{0.12}_{0.11}$&$0.98^{0.1}_{0.093}$&$17.0^{1.5}_{1.5}$&$52.0^{2.6}_{3.2}$\\[1.5pt]\hline

        L&All&$0.19^{0.02}_{0.017}$&$0.3^{0.027}_{0.029}$&$14.0^{1.2}_{1.2}$&$25.0^{2.0}_{1.8}$&$1.6^{0.12}_{0.12}$&$3.4^{0.25}_{0.24}$&$4.0^{0.36}_{0.35}$&$2.7^{0.26}_{0.23}$&$1.5^{0.11}_{0.099}$&$-3.2^{0.2}_{0.23}$&$1.9^{0.19}_{0.2}$&$0.7^{0.076}_{0.067}$&$17.0^{1.3}_{1.2}$&$24.0^{1.8}_{2.0}$\\[1.5pt]
        &S1&$0.19^{0.024}_{0.018}$&$0.3^{0.031}_{0.027}$&$14.0^{1.2}_{0.88}$&$25.0^{1.6}_{1.9}$&$1.5^{0.14}_{0.13}$&$3.4^{0.21}_{0.22}$&$4.2^{0.41}_{0.4}$&$2.6^{0.34}_{0.27}$&$1.5^{0.073}_{0.078}$&$-3.0^{0.22}_{0.25}$&$1.8^{0.22}_{0.27}$&$0.73^{0.087}_{0.085}$&$17.0^{1.3}_{1.5}$&$24.0^{2.4}_{2.5}$\\[1.5pt]
        &S2&$0.2^{0.019}_{0.021}$&$0.3^{0.027}_{0.035}$&$14.0^{1.3}_{1.2}$&$25.0^{1.9}_{2.1}$&$1.6^{0.11}_{0.094}$&$3.4^{0.25}_{0.23}$&$4.0^{0.36}_{0.28}$&$2.7^{0.24}_{0.16}$&$1.5^{0.11}_{0.11}$&$-2.9^{0.18}_{0.19}$&$1.9^{0.2}_{0.18}$&$0.68^{0.067}_{0.056}$&$17.0^{1.2}_{1.2}$&$24.0^{1.8}_{2.0}$\\[1.5pt]
        &S3&$0.2^{0.024}_{0.019}$&$0.29^{0.022}_{0.023}$&$14.0^{1.2}_{1.2}$&$25.0^{2.2}_{1.9}$&$1.6^{0.12}_{0.12}$&$3.4^{0.21}_{0.22}$&$4.0^{0.35}_{0.39}$&$2.7^{0.26}_{0.22}$&$1.5^{0.11}_{0.093}$&$-2.9^{0.18}_{0.26}$&$1.9^{0.21}_{0.21}$&$0.7^{0.063}_{0.053}$&$16.0^{1.4}_{1.1}$&$24.0^{1.8}_{1.8}$\\[1.5pt]
        &S4&$0.19^{0.015}_{0.017}$&$0.29^{0.033}_{0.029}$&$15.0^{1.1}_{1.2}$&$25.0^{2.8}_{2.1}$&$1.6^{0.12}_{0.095}$&$3.4^{0.31}_{0.27}$&$3.9^{0.32}_{0.26}$&$2.8^{0.2}_{0.23}$&$1.5^{0.12}_{0.13}$&$-2.9^{0.22}_{0.18}$&$1.9^{0.14}_{0.16}$&$0.69^{0.08}_{0.082}$&$17.0^{1.4}_{1.1}$&$24.0^{1.7}_{1.9}$\\[1.5pt]
        &S5&$0.19^{0.016}_{0.016}$&$0.3^{0.022}_{0.031}$&$14.0^{1.3}_{1.4}$&$25.0^{1.7}_{1.5}$&$1.6^{0.13}_{0.11}$&$3.4^{0.27}_{0.22}$&$4.0^{0.34}_{0.33}$&$2.8^{0.2}_{0.26}$&$1.5^{0.11}_{0.11}$&$-2.9^{0.2}_{0.2}$&$1.9^{0.14}_{0.18}$&$0.68^{0.073}_{0.062}$&$17.0^{1.1}_{0.88}$&$24.0^{1.5}_{1.6}$\\[1.5pt]\hline

        M&All&$0.23^{0.024}_{0.025}$&$0.8^{0.073}_{0.064}$&$30.0^{2.6}_{3.2}$&$18.0^{1.9}_{1.6}$&$1.6^{0.14}_{0.13}$&$3.4^{0.27}_{0.27}$&$3.9^{0.38}_{0.33}$&$2.2^{0.2}_{0.21}$&$1.1^{0.089}_{0.089}$&$-3.9^{0.32}_{0.31}$&$3.3^{0.29}_{0.29}$&$0.83^{0.077}_{0.078}$&$15.0^{1.0}_{1.1}$&$29.0^{2.3}_{2.3}$\\[1.5pt]

        &S1&$0.23^{0.026}_{0.032}$&$0.82^{0.066}_{0.07}$&$30.0^{2.5}_{3.0}$&$19.0^{2.1}_{1.6}$&$1.7^{0.13}_{0.14}$&$3.3^{0.3}_{0.3}$&$3.8^{0.39}_{0.29}$&$2.2^{0.23}_{0.25}$&$1.1^{0.12}_{0.089}$&$-3.6^{0.36}_{0.32}$&$3.4^{0.25}_{0.3}$&$0.84^{0.071}_{0.073}$&$15.0^{0.94}_{0.85}$&$28.0^{2.6}_{2.7}$\\[1.5pt]
        &S2&$0.22^{0.028}_{0.024}$&$0.8^{0.073}_{0.063}$&$30.0^{2.7}_{2.4}$&$18.0^{1.7}_{1.5}$&$1.6^{0.12}_{0.09}$&$3.4^{0.23}_{0.22}$&$3.8^{0.3}_{0.2}$&$2.2^{0.18}_{0.16}$&$1.1^{0.054}_{0.063}$&$-3.6^{0.29}_{0.28}$&$3.4^{0.34}_{0.36}$&$0.82^{0.055}_{0.062}$&$15.0^{1.0}_{1.1}$&$29.0^{2.0}_{2.4}$\\[1.5pt]
        &S3&$0.23^{0.021}_{0.023}$&$0.81^{0.093}_{0.072}$&$30.0^{2.3}_{2.1}$&$18.0^{1.7}_{1.8}$&$1.7^{0.19}_{0.16}$&$3.4^{0.26}_{0.24}$&$3.9^{0.29}_{0.35}$&$2.2^{0.16}_{0.2}$&$1.1^{0.097}_{0.11}$&$-3.6^{0.34}_{0.36}$&$3.2^{0.28}_{0.23}$&$0.8^{0.08}_{0.091}$&$15.0^{1.3}_{1.3}$&$29.0^{2.3}_{2.2}$\\[1.5pt]
        &S4&$0.23^{0.026}_{0.02}$&$0.79^{0.06}_{0.075}$&$30.0^{2.8}_{3.9}$&$18.0^{1.7}_{1.5}$&$1.6^{0.11}_{0.13}$&$3.3^{0.28}_{0.29}$&$3.8^{0.42}_{0.33}$&$2.3^{0.19}_{0.23}$&$1.1^{0.1}_{0.09}$&$-3.6^{0.24}_{0.26}$&$3.3^{0.27}_{0.28}$&$0.83^{0.088}_{0.087}$&$15.0^{0.69}_{0.99}$&$29.0^{2.5}_{2.2}$\\[1.5pt]
        &S5&$0.24^{0.025}_{0.025}$&$0.8^{0.052}_{0.045}$&$30.0^{2.8}_{3.9}$&$18.0^{1.7}_{1.6}$&$1.6^{0.15}_{0.15}$&$3.3^{0.29}_{0.28}$&$3.9^{0.47}_{0.39}$&$2.2^{0.24}_{0.24}$&$1.0^{0.086}_{0.071}$&$-3.7^{0.41}_{0.41}$&$3.3^{0.27}_{0.31}$&$0.83^{0.08}_{0.082}$&$15.0^{1.2}_{1.3}$&$29.0^{1.9}_{2.2}$\\[1.5pt]\hline

        N&All&$0.57^{0.051}_{0.051}$&$0.6^{0.057}_{0.059}$&$230.0^{22.0}_{20.0}$&$24.0^{2.0}_{2.3}$&$1.3^{0.1}_{0.1}$&$3.5^{0.28}_{0.3}$&$3.0^{0.24}_{0.25}$&$3.3^{0.29}_{0.28}$&$1.1^{0.091}_{0.084}$&$-3.1^{0.091}_{0.11}$&$3.5^{0.32}_{0.32}$&$0.31^{0.03}_{0.025}$&$11.0^{0.9}_{0.86}$&$37.0^{2.2}_{2.3}$\\[1.5pt]

        &S1&$0.57^{0.058}_{0.062}$&$0.61^{0.052}_{0.074}$&$230.0^{19.0}_{20.0}$&$24.0^{2.1}_{2.0}$&$1.3^{0.13}_{0.12}$&$3.6^{0.31}_{0.33}$&$2.9^{0.24}_{0.21}$&$3.3^{0.32}_{0.32}$&$1.2^{0.098}_{0.11}$&$-2.9^{0.082}_{0.1}$&$3.5^{0.32}_{0.37}$&$0.31^{0.032}_{0.025}$&$11.0^{0.89}_{0.91}$&$37.0^{2.6}_{2.6}$\\[1.5pt]
        &S2&$0.57^{0.071}_{0.057}$&$0.61^{0.048}_{0.062}$&$230.0^{18.0}_{17.0}$&$24.0^{1.5}_{2.0}$&$1.3^{0.084}_{0.098}$&$3.5^{0.36}_{0.29}$&$3.0^{0.24}_{0.23}$&$3.3^{0.3}_{0.25}$&$1.1^{0.085}_{0.067}$&$-2.9^{0.093}_{0.092}$&$3.4^{0.38}_{0.32}$&$0.31^{0.028}_{0.023}$&$11.0^{1.1}_{0.9}$&$37.0^{2.3}_{2.3}$\\[1.5pt]
        &S3&$0.56^{0.035}_{0.039}$&$0.59^{0.051}_{0.067}$&$230.0^{20.0}_{19.0}$&$24.0^{2.5}_{2.0}$&$1.3^{0.099}_{0.09}$&$3.6^{0.31}_{0.35}$&$3.0^{0.3}_{0.28}$&$3.2^{0.26}_{0.23}$&$1.1^{0.084}_{0.083}$&$-2.9^{0.08}_{0.12}$&$3.5^{0.27}_{0.24}$&$0.31^{0.025}_{0.021}$&$11.0^{0.92}_{0.92}$&$38.0^{2.1}_{1.8}$\\[1.5pt]
        &S4&$0.57^{0.051}_{0.054}$&$0.6^{0.066}_{0.065}$&$240.0^{23.0}_{25.0}$&$24.0^{2.1}_{3.5}$&$1.3^{0.1}_{0.12}$&$3.5^{0.22}_{0.32}$&$3.0^{0.23}_{0.21}$&$3.3^{0.23}_{0.23}$&$1.1^{0.1}_{0.077}$&$-2.9^{0.11}_{0.12}$&$3.5^{0.26}_{0.33}$&$0.31^{0.033}_{0.026}$&$10.0^{0.87}_{0.94}$&$37.0^{2.2}_{2.1}$\\[1.5pt]
        &S5&$0.56^{0.05}_{0.047}$&$0.59^{0.054}_{0.043}$&$230.0^{25.0}_{20.0}$&$25.0^{2.0}_{2.1}$&$1.3^{0.093}_{0.071}$&$3.5^{0.23}_{0.23}$&$3.0^{0.2}_{0.24}$&$3.3^{0.3}_{0.4}$&$1.1^{0.075}_{0.079}$&$-2.9^{0.098}_{0.095}$&$3.4^{0.36}_{0.34}$&$0.32^{0.028}_{0.031}$&$11.0^{0.68}_{0.66}$&$37.0^{1.9}_{2.4}$\\[1.5pt]\hline

        \end{tabular*}
    \tablefoot{Median parameters of the final models with 1$\sigma$ errors. All in Band is the polychromatic fit. When the Seed is given as All, it is the results from the combined probability distributions of the different seeds.}
    \label{tab:results}
\end{table}
\end{landscape}
\begin{landscape}

\setcounter{table}{0}

\begin{table}

\normalsize
    \caption{Continued - Model Results}
    \centering
    \begin{tabular*}{\linewidth}{@{\extracolsep{\fill}} ccccccccccccccccc}\hline
        Band&Seed&$p_g$&$n_c$&$N_p$&$f_w$&$\alpha_w$&$\ln f_v$&$\ln f_c$&$s_c$&$T_{bw}$&$T_{bd}$&$r_c$&$BB_\mathrm{hot}$&$V_0$&\\[1.5pt]\hline
        &&\multicolumn{3}{c}{$log_{10}$($\frac{\mathrm{g}}{\mathrm{ cm}^2 \mathrm{mas}}$)}&&&&&&K&K&mas&$F_\mathrm{AGN}\left(\lambda_\mathrm{min}\right)$&&\\[1.5pt]\hline\hline
        All&SP1&$0.16^{0.01}_{0.01}$&$-3.9^{0.23}_{0.3}$&$18000.0^{1400.0}_{1400.0}$&$-1.7^{0.14}_{0.11}$&$1.7^{0.13}_{0.15}$&$-12.0^{0.91}_{0.84}$&$-9.7^{0.49}_{0.71}$&$0.71^{0.08}_{0.06}$&$280.0^{26.0}_{26.0}$&$160.0^{12.0}_{8.5}$&$1.1^{0.11}_{0.096}$&$0.32^{0.03}_{0.04}$&$0.65^{0.063}_{0.073}$\\[1.5pt]\hline

        K&All&$0.087^{0.0071}_{0.0075}$&$-3.9^{0.17}_{0.18}$&$17000.0^{1200.0}_{1300.0}$&$-1.2^{0.077}_{0.088}$&$1.3^{0.1}_{0.084}$&$-9.7^{0.4}_{0.57}$&$-9.8^{0.47}_{0.62}$&$0.57^{0.044}_{0.046}$&$120.0^{8.7}_{9.8}$&$210.0^{15.0}_{15.0}$&$0.84^{0.081}_{0.053}$&$0.46^{0.04}_{0.037}$&$0.55^{0.033}_{0.03}$\\[1.5pt]
        &S1&$0.089^{0.0049}_{0.0052}$&$-3.9^{0.11}_{0.15}$&$17000.0^{1000.0}_{1000.0}$&$-1.2^{0.043}_{0.053}$&$1.3^{0.089}_{0.074}$&$-9.5^{0.29}_{0.33}$&$-9.6^{0.32}_{0.4}$&$0.57^{0.034}_{0.03}$&$120.0^{8.0}_{7.6}$&$210.0^{10.0}_{11.0}$&$0.84^{0.06}_{0.043}$&$0.46^{0.026}_{0.019}$&$0.54^{0.025}_{0.024}$\\[1.5pt]
        &S2&$0.087^{0.0055}_{0.0051}$&$-3.9^{0.15}_{0.16}$&$17000.0^{940.0}_{1200.0}$&$-1.2^{0.058}_{0.072}$&$1.3^{0.078}_{0.066}$&$-9.7^{0.36}_{0.59}$&$-9.8^{0.58}_{0.69}$&$0.58^{0.038}_{0.038}$&$110.0^{8.8}_{10.0}$&$210.0^{16.0}_{9.5}$&$0.84^{0.076}_{0.052}$&$0.46^{0.031}_{0.03}$&$0.54^{0.028}_{0.026}$\\[1.5pt]
        &S3&$0.085^{0.0091}_{0.011}$&$-3.9^{0.18}_{0.19}$&$17000.0^{1300.0}_{1100.0}$&$-1.2^{0.075}_{0.088}$&$1.3^{0.12}_{0.086}$&$-9.9^{0.49}_{0.65}$&$-9.9^{0.43}_{0.62}$&$0.56^{0.058}_{0.048}$&$120.0^{7.4}_{10.0}$&$200.0^{14.0}_{18.0}$&$0.83^{0.083}_{0.056}$&$0.45^{0.04}_{0.049}$&$0.55^{0.046}_{0.036}$\\[1.5pt]
        &S4&$0.088^{0.0075}_{0.0087}$&$-3.9^{0.19}_{0.18}$&$18000.0^{990.0}_{1500.0}$&$-1.2^{0.1}_{0.077}$&$1.3^{0.11}_{0.086}$&$-9.7^{0.46}_{0.54}$&$-9.9^{0.5}_{0.76}$&$0.56^{0.051}_{0.052}$&$120.0^{9.1}_{9.2}$&$210.0^{25.0}_{23.0}$&$0.84^{0.1}_{0.074}$&$0.47^{0.048}_{0.047}$&$0.55^{0.038}_{0.033}$\\[1.5pt]
        &S5&$0.088^{0.0079}_{0.0068}$&$-3.9^{0.17}_{0.18}$&$17000.0^{1200.0}_{1300.0}$&$-1.2^{0.097}_{0.11}$&$1.3^{0.12}_{0.11}$&$-9.7^{0.46}_{0.53}$&$-9.7^{0.45}_{0.58}$&$0.57^{0.035}_{0.051}$&$120.0^{8.9}_{10.0}$&$210.0^{13.0}_{15.0}$&$0.83^{0.078}_{0.043}$&$0.46^{0.048}_{0.046}$&$0.55^{0.033}_{0.03}$\\[1.5pt]\hline

        L&All&$0.13^{0.013}_{0.013}$&$-3.2^{0.19}_{0.19}$&$14000.0^{890.0}_{890.0}$&$-1.1^{0.091}_{0.088}$&$1.2^{0.096}_{0.1}$&$-9.7^{0.46}_{0.68}$&$-9.6^{0.43}_{0.54}$&$0.72^{0.059}_{0.061}$&$210.0^{20.0}_{23.0}$&$180.0^{16.0}_{14.0}$&$0.49^{0.039}_{0.034}$&$0.05^{0.0046}_{0.004}$&$0.75^{0.058}_{0.05}$\\[1.5pt]

        &S1&$0.13^{0.013}_{0.013}$&$-3.2^{0.23}_{0.18}$&$14000.0^{1600.0}_{1100.0}$&$-1.1^{0.11}_{0.11}$&$1.2^{0.096}_{0.093}$&$-9.7^{0.49}_{0.57}$&$-9.6^{0.42}_{0.69}$&$0.7^{0.065}_{0.056}$&$210.0^{23.0}_{29.0}$&$180.0^{14.0}_{12.0}$&$0.5^{0.042}_{0.037}$&$0.05^{0.0041}_{0.0036}$&$0.76^{0.06}_{0.048}$\\[1.5pt]
        &S2&$0.13^{0.012}_{0.014}$&$-3.1^{0.14}_{0.18}$&$14000.0^{790.0}_{1100.0}$&$-1.1^{0.075}_{0.071}$&$1.3^{0.096}_{0.12}$&$-9.7^{0.5}_{0.61}$&$-9.7^{0.47}_{0.54}$&$0.72^{0.046}_{0.069}$&$220.0^{18.0}_{21.0}$&$170.0^{17.0}_{12.0}$&$0.49^{0.037}_{0.03}$&$0.05^{0.0036}_{0.0035}$&$0.76^{0.056}_{0.06}$\\[1.5pt]
        &S3&$0.13^{0.011}_{0.013}$&$-3.2^{0.19}_{0.14}$&$14000.0^{970.0}_{620.0}$&$-1.1^{0.11}_{0.098}$&$1.3^{0.076}_{0.085}$&$-9.8^{0.41}_{0.82}$&$-9.6^{0.37}_{0.62}$&$0.73^{0.059}_{0.069}$&$220.0^{23.0}_{18.0}$&$180.0^{17.0}_{14.0}$&$0.5^{0.037}_{0.033}$&$0.049^{0.004}_{0.0037}$&$0.76^{0.062}_{0.053}$\\[1.5pt]
        &S4&$0.13^{0.012}_{0.011}$&$-3.1^{0.19}_{0.2}$&$14000.0^{980.0}_{650.0}$&$-1.1^{0.074}_{0.076}$&$1.2^{0.14}_{0.11}$&$-9.6^{0.39}_{0.49}$&$-9.7^{0.5}_{0.42}$&$0.71^{0.062}_{0.066}$&$210.0^{20.0}_{19.0}$&$180.0^{23.0}_{17.0}$&$0.49^{0.037}_{0.045}$&$0.051^{0.0039}_{0.0051}$&$0.74^{0.054}_{0.045}$\\[1.5pt]
        &S5&$0.13^{0.012}_{0.013}$&$-3.2^{0.21}_{0.22}$&$14000.0^{770.0}_{560.0}$&$-1.1^{0.097}_{0.077}$&$1.2^{0.082}_{0.075}$&$-9.8^{0.51}_{0.7}$&$-9.6^{0.4}_{0.38}$&$0.72^{0.055}_{0.055}$&$220.0^{16.0}_{24.0}$&$180.0^{16.0}_{16.0}$&$0.49^{0.046}_{0.031}$&$0.051^{0.0049}_{0.0055}$&$0.74^{0.048}_{0.043}$\\[1.5pt]\hline

        M&All&$0.14^{0.013}_{0.014}$&$-3.5^{0.19}_{0.2}$&$11000.0^{1200.0}_{1200.0}$&$-0.77^{0.067}_{0.071}$&$0.38^{0.033}_{0.031}$&$-9.8^{0.46}_{0.76}$&$-9.5^{0.33}_{0.43}$&$0.86^{0.088}_{0.073}$&$84.0^{8.3}_{7.6}$&$180.0^{17.0}_{17.0}$&$0.74^{0.069}_{0.061}$&$0.087^{0.0076}_{0.0074}$&$0.56^{0.035}_{0.035}$\\[1.5pt]
        &S1&$0.14^{0.015}_{0.016}$&$-3.5^{0.23}_{0.22}$&$11000.0^{1500.0}_{1400.0}$&$-0.77^{0.073}_{0.081}$&$0.38^{0.044}_{0.035}$&$-9.9^{0.6}_{0.82}$&$-9.6^{0.35}_{0.42}$&$0.86^{0.085}_{0.068}$&$84.0^{9.1}_{7.5}$&$190.0^{26.0}_{20.0}$&$0.75^{0.067}_{0.06}$&$0.089^{0.0065}_{0.0085}$&$0.56^{0.036}_{0.036}$\\[1.5pt]
        &S2&$0.14^{0.015}_{0.013}$&$-3.5^{0.17}_{0.17}$&$11000.0^{1200.0}_{1000.0}$&$-0.76^{0.05}_{0.07}$&$0.39^{0.026}_{0.025}$&$-9.8^{0.42}_{0.8}$&$-9.5^{0.35}_{0.48}$&$0.86^{0.086}_{0.052}$&$86.0^{10.0}_{11.0}$&$180.0^{15.0}_{21.0}$&$0.76^{0.072}_{0.056}$&$0.086^{0.0075}_{0.007}$&$0.55^{0.045}_{0.034}$\\[1.5pt]
        &S3&$0.13^{0.017}_{0.013}$&$-3.5^{0.19}_{0.22}$&$11000.0^{1100.0}_{960.0}$&$-0.77^{0.064}_{0.063}$&$0.38^{0.032}_{0.031}$&$-9.8^{0.43}_{0.66}$&$-9.5^{0.32}_{0.44}$&$0.83^{0.083}_{0.083}$&$85.0^{6.1}_{7.1}$&$180.0^{15.0}_{17.0}$&$0.74^{0.057}_{0.066}$&$0.086^{0.0069}_{0.0066}$&$0.55^{0.034}_{0.031}$\\[1.5pt]
        &S4&$0.13^{0.014}_{0.013}$&$-3.5^{0.19}_{0.14}$&$11000.0^{1200.0}_{1000.0}$&$-0.76^{0.058}_{0.056}$&$0.38^{0.032}_{0.033}$&$-9.8^{0.54}_{0.66}$&$-9.5^{0.31}_{0.57}$&$0.87^{0.065}_{0.064}$&$83.0^{6.0}_{7.9}$&$180.0^{17.0}_{15.0}$&$0.74^{0.068}_{0.068}$&$0.088^{0.0071}_{0.009}$&$0.56^{0.032}_{0.038}$\\[1.5pt]
        &S5&$0.14^{0.0091}_{0.011}$&$-3.5^{0.2}_{0.17}$&$11000.0^{930.0}_{1300.0}$&$-0.78^{0.086}_{0.1}$&$0.38^{0.034}_{0.031}$&$-9.7^{0.39}_{0.6}$&$-9.5^{0.34}_{0.34}$&$0.87^{0.11}_{0.071}$&$84.0^{9.5}_{5.9}$&$180.0^{15.0}_{13.0}$&$0.73^{0.066}_{0.054}$&$0.087^{0.0083}_{0.0076}$&$0.55^{0.033}_{0.036}$\\[1.5pt]\hline

        N&All&$0.15^{0.013}_{0.014}$&$-2.8^{0.23}_{0.21}$&$16000.0^{1700.0}_{1600.0}$&$-1.2^{0.1}_{0.099}$&$0.87^{0.067}_{0.062}$&$-9.6^{0.37}_{0.51}$&$-9.8^{0.47}_{0.59}$&$0.66^{0.065}_{0.06}$&$300.0^{18.0}_{20.0}$&$160.0^{11.0}_{12.0}$&$1.1^{0.076}_{0.052}$&$0.21^{0.02}_{0.02}$&$0.66^{0.056}_{0.054}$\\[1.5pt]
        &S1&$0.15^{0.013}_{0.012}$&$-2.8^{0.3}_{0.22}$&$16000.0^{1800.0}_{1700.0}$&$-1.2^{0.1}_{0.094}$&$0.86^{0.056}_{0.046}$&$-9.6^{0.4}_{0.52}$&$-9.8^{0.49}_{0.68}$&$0.68^{0.062}_{0.064}$&$300.0^{15.0}_{17.0}$&$170.0^{11.0}_{12.0}$&$1.1^{0.086}_{0.066}$&$0.21^{0.019}_{0.016}$&$0.66^{0.061}_{0.052}$\\[1.5pt]
        &S2&$0.15^{0.013}_{0.016}$&$-2.8^{0.17}_{0.16}$&$16000.0^{2100.0}_{1300.0}$&$-1.2^{0.12}_{0.098}$&$0.87^{0.052}_{0.083}$&$-9.6^{0.35}_{0.61}$&$-9.8^{0.46}_{0.56}$&$0.66^{0.064}_{0.062}$&$290.0^{20.0}_{14.0}$&$160.0^{11.0}_{13.0}$&$1.1^{0.067}_{0.053}$&$0.21^{0.02}_{0.019}$&$0.66^{0.05}_{0.057}$\\[1.5pt]
        &S3&$0.16^{0.013}_{0.014}$&$-2.9^{0.23}_{0.17}$&$16000.0^{1500.0}_{1700.0}$&$-1.2^{0.086}_{0.08}$&$0.85^{0.069}_{0.044}$&$-9.6^{0.37}_{0.54}$&$-9.9^{0.49}_{0.58}$&$0.65^{0.044}_{0.044}$&$300.0^{16.0}_{17.0}$&$160.0^{11.0}_{12.0}$&$1.1^{0.08}_{0.049}$&$0.21^{0.02}_{0.024}$&$0.65^{0.055}_{0.046}$\\[1.5pt]
        &S4&$0.16^{0.014}_{0.013}$&$-2.8^{0.28}_{0.24}$&$17000.0^{1800.0}_{1800.0}$&$-1.2^{0.11}_{0.097}$&$0.88^{0.084}_{0.068}$&$-9.7^{0.39}_{0.37}$&$-9.9^{0.49}_{0.57}$&$0.64^{0.087}_{0.068}$&$290.0^{20.0}_{26.0}$&$160.0^{11.0}_{13.0}$&$1.1^{0.066}_{0.044}$&$0.21^{0.024}_{0.024}$&$0.64^{0.073}_{0.059}$\\[1.5pt]
        &S5&$0.15^{0.013}_{0.013}$&$-2.8^{0.22}_{0.22}$&$16000.0^{1400.0}_{1300.0}$&$-1.2^{0.072}_{0.1}$&$0.88^{0.072}_{0.081}$&$-9.5^{0.35}_{0.41}$&$-9.8^{0.46}_{0.51}$&$0.67^{0.059}_{0.058}$&$300.0^{17.0}_{22.0}$&$160.0^{9.9}_{9.5}$&$1.0^{0.068}_{0.059}$&$0.21^{0.018}_{0.015}$&$0.67^{0.041}_{0.049}$\\[1.5pt]\hline

    \end{tabular*}
\end{table}
\end{landscape}

\begin{figure*}
    \centering
    \includegraphics[width=0.9\textwidth]{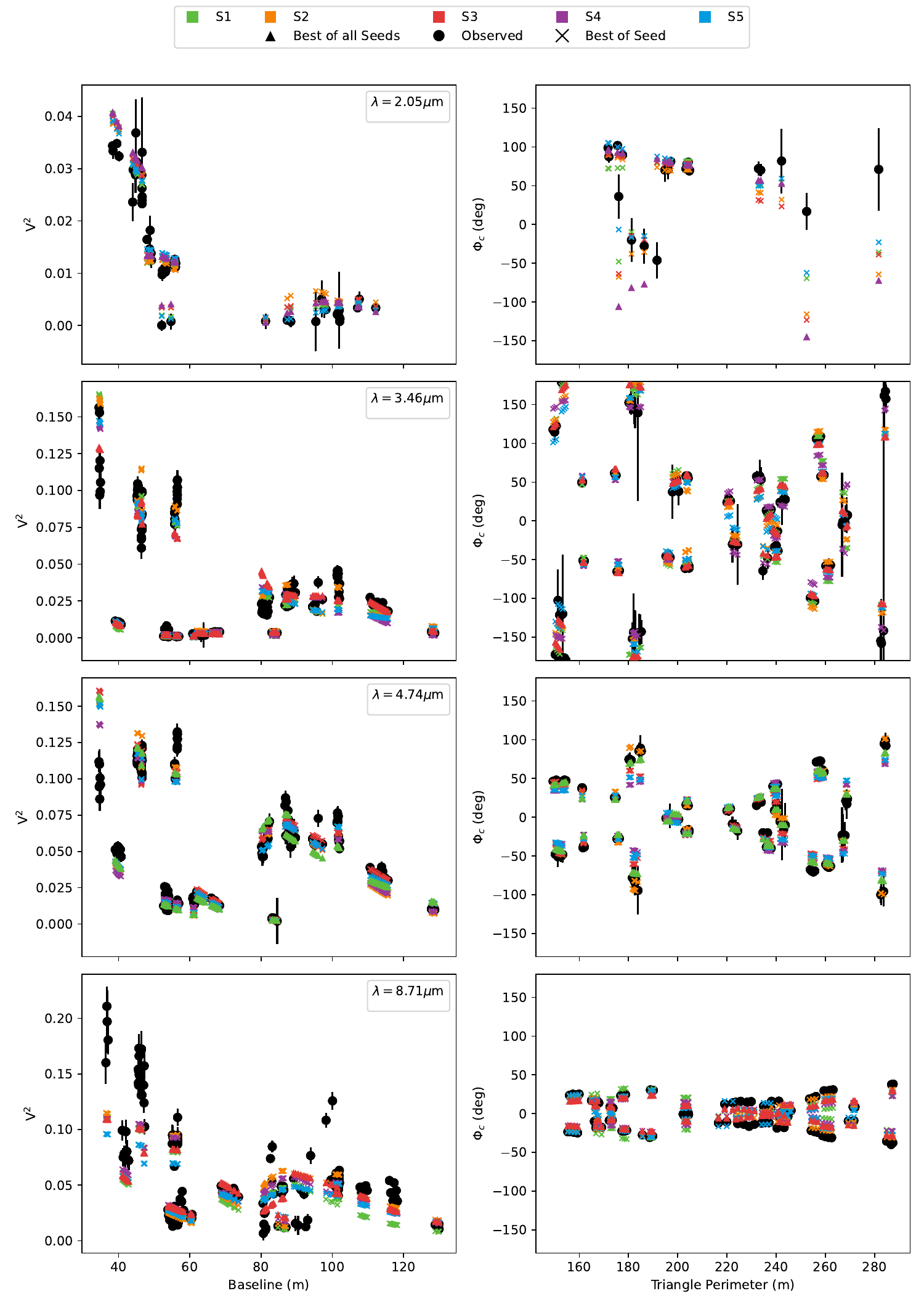}
    \caption{V$^2$ and $\Phi_c$ from the best models in Sect.\,\ref{sec:finalind} at one wavelength per band for each seed compared to the observed data.}
    \label{fig:observables_final}
\end{figure*}

\begin{figure*}
    \centering
    \includegraphics[width=\textwidth]{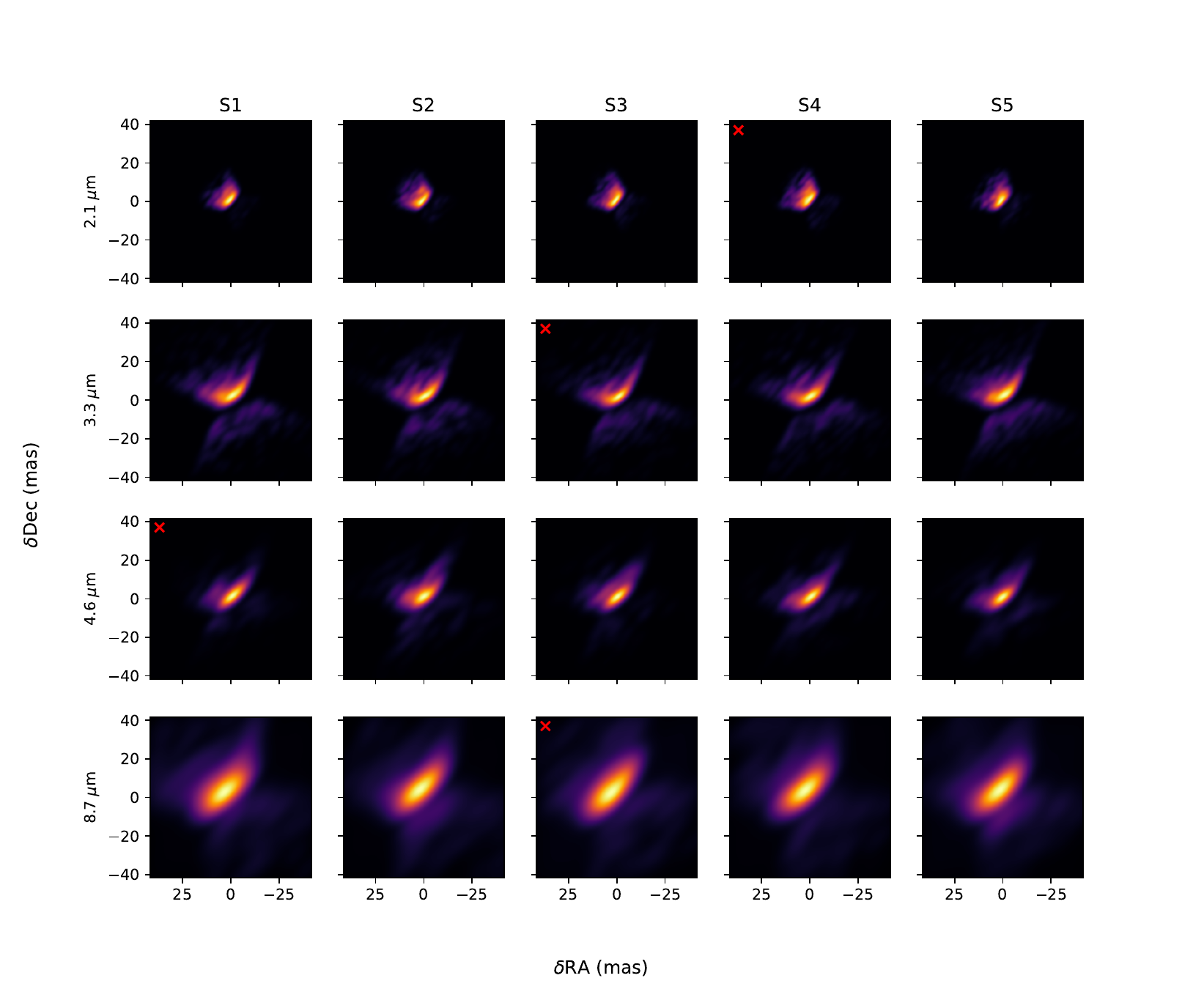}
    \caption{Best model for each band (row) and each seed (column) for the final band-by-band fit as determined by the maximum likelihood. The best seed for each band is marked with a red x at the top left corner. The K-,L-,M-,N-bands are evaluated at 2.1\,$\mu$m, 3.5\,$\mu$m, 4.7\,$\mu$m, and 8.5\,$\mu$m, respectively. All images are normalised and given a 0.6 power colour scaling to match that of \citet{gamez_rosas_thermal_2022} for easy comparison.}
    \label{fig:bbb_final}
\end{figure*}

\begin{figure*}
    \centering
    \includegraphics[width=\textwidth]{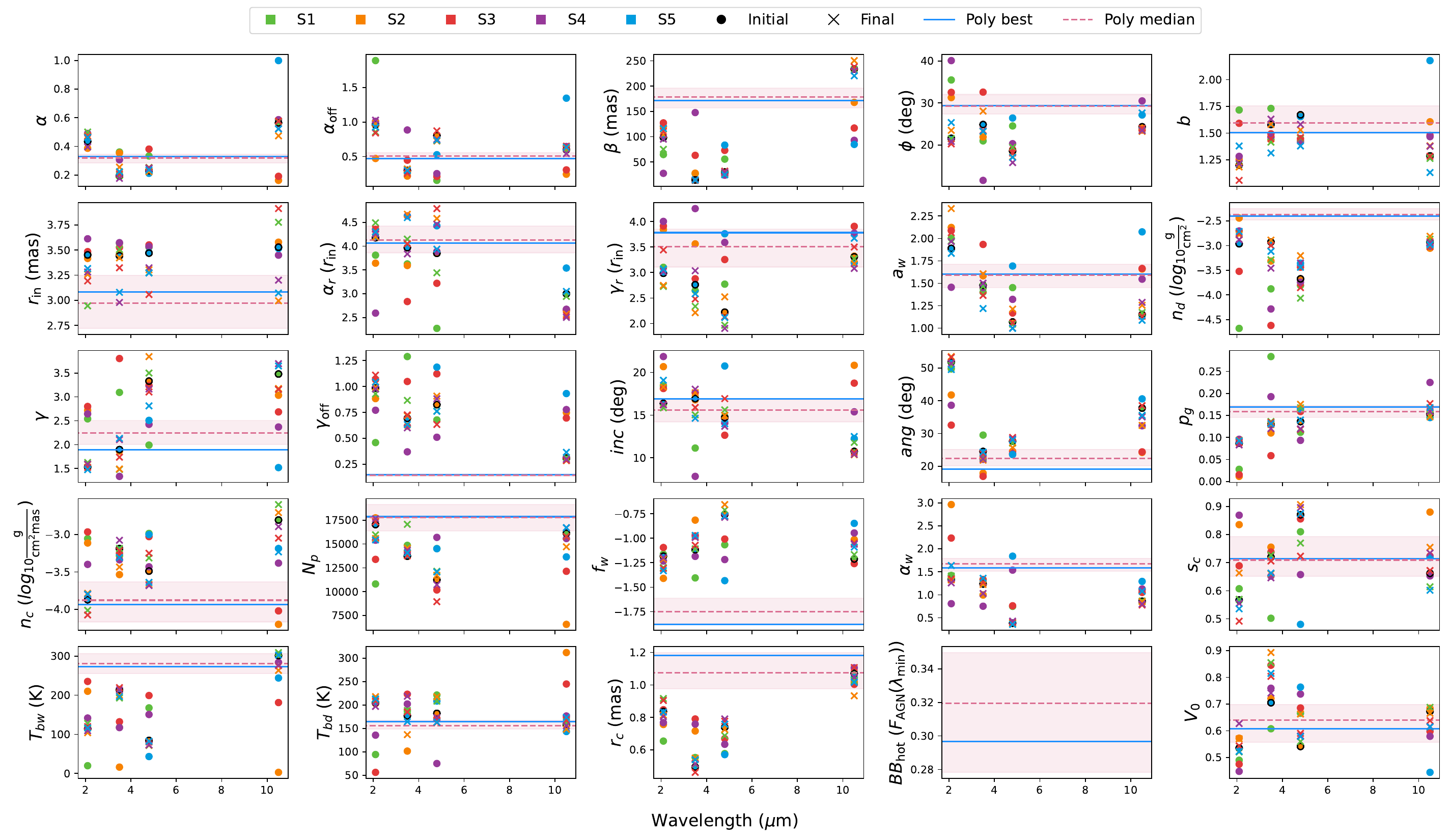}
    \caption{Parameters of the best model for each band and each seed (represented by colours as given in legend) for the initial band-by-band fit (coloured circles) and final band-by-band fit (coloured crosses) as determined by the maximum likelihood. The seed used as the starting position for the final fit in each band is given a black border. Overplotted as lines are the best and median fit for the polychromatic model. The shaded region is the 1$\sigma$ errors on the median fit parameters.}
    \label{fig:results_initial}
\end{figure*}

\begin{figure*}
    \centering
    \includegraphics[width=0.85\textwidth]{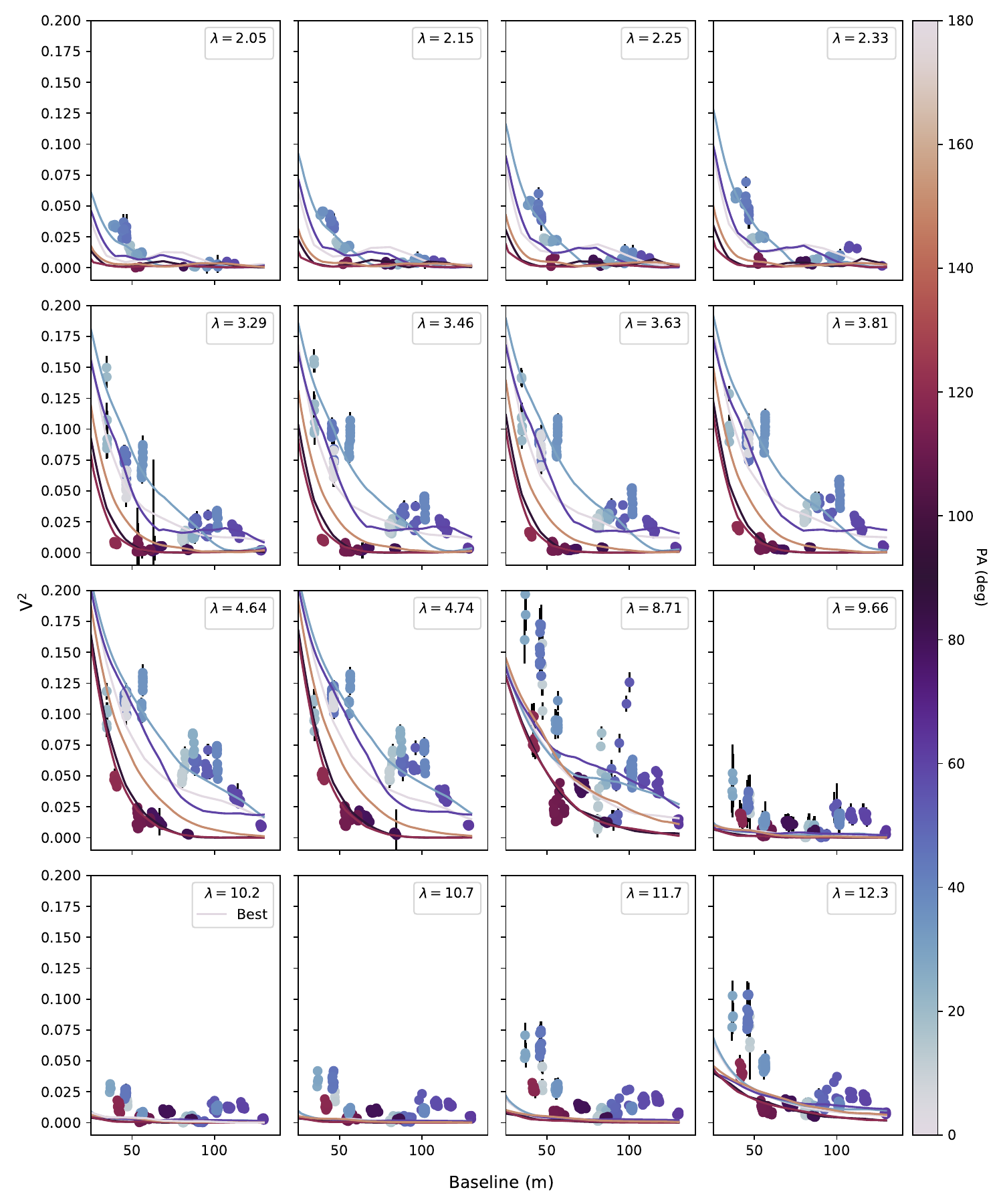}
    \caption{V$^2$ of the polychromatic model. {The colours represent the PA on sky of the telescope pair used for that point. The colour scheme is continuous and cyclical about $\pi$, due to the fact that a V$^2$ measurement at PA $=x$ and $x+\pi$ are equivalent. The line is the best polychromatic model by maximum likelihood evaluated at different PA in $\frac{\pi}{6}$ intervals.}}
    \label{fig:polyv2}
\end{figure*}

\begin{figure*}
    \centering
    \includegraphics[width=0.85\textwidth]{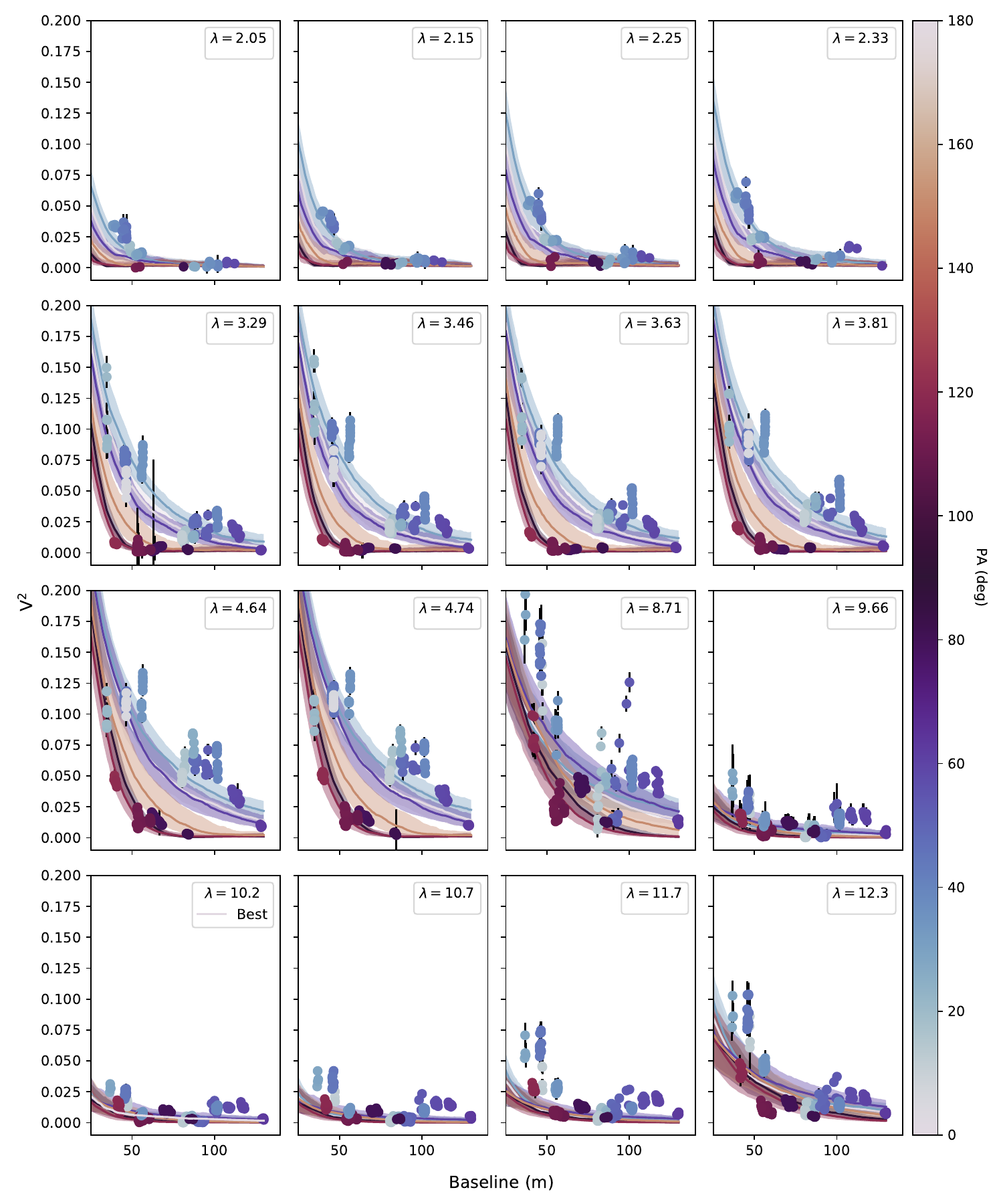}
    \caption{V$^2$ of the polychromatic model. {The colours represent the PA on sky of the telescope pair used for that point. The colour scheme is continuous and cyclical about $\pi$, due to the fact that a V$^2$ measurement at PA $=x$ and $x+\pi$ are equivalent. The line is the median fit of the polychromatic model with 1$\,\sigma$ errors as the shaded region evaluated at different PA in $\frac{\pi}{6}$ intervals.}}
    \label{fig:polyv2med}
\end{figure*}

\begin{figure*}
    \centering
    \includegraphics[width=\textwidth]{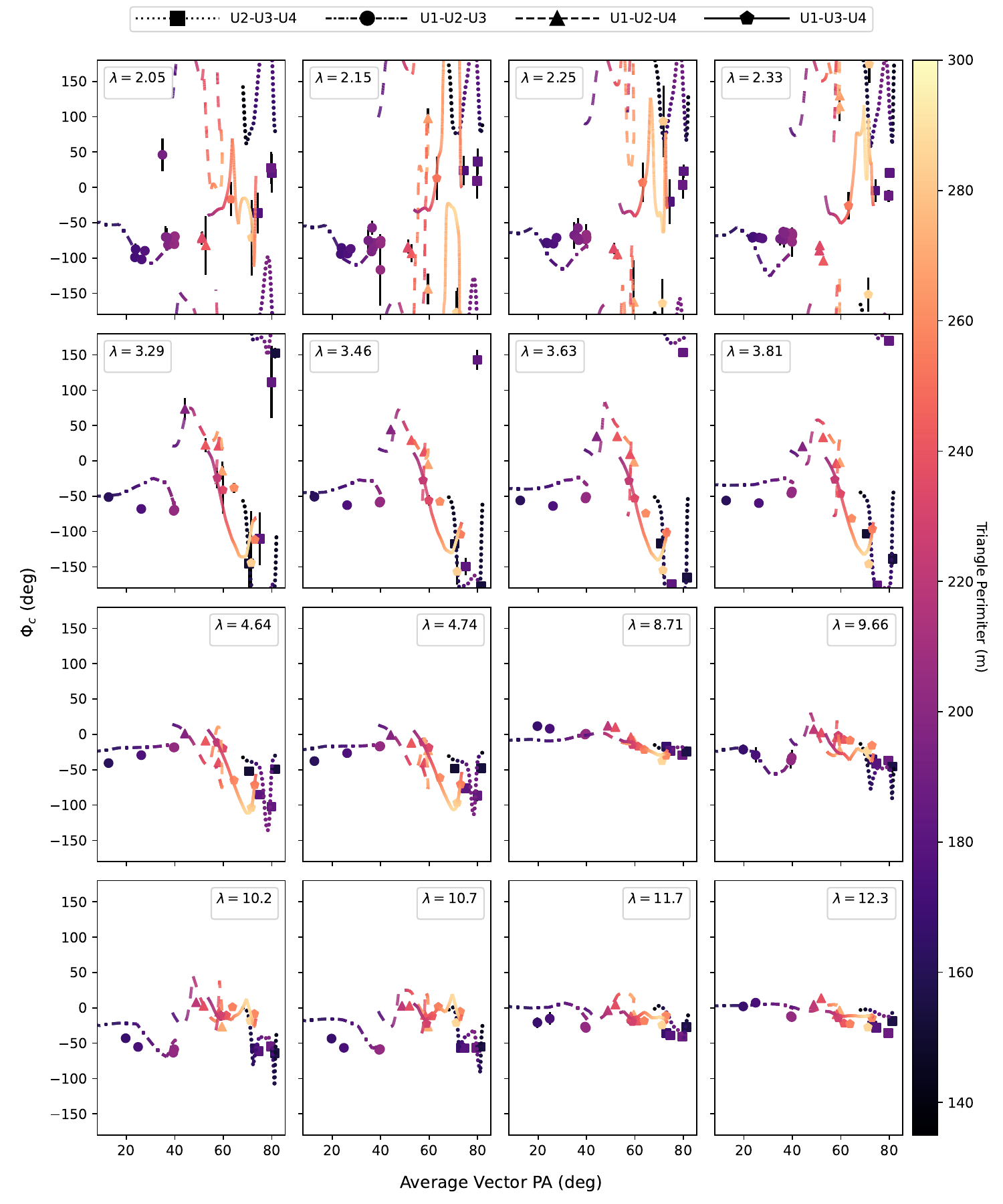}
    \caption{Closure phase of the polychromatic model per wavelength compared to observations. The closure phases for the best fitting model by maximum likelihood were calculated for each of the four telescope triplets at all accessible \textit{uv} positions for the UTs where NGC\,1068 has a minimum altitude of 35$^\circ$. The average vector PA is the average of the PA for each of the three baselines in the triplet ensuring the telescopes are in ascending numerical order. E.g. the average PA for U1-U2-U3 is the is average of the PA of the baselines U1-U2, U2-U3, and U1-U3 {not U3-U1}.  For visual purposes, we have averaged the different BCD positions for the MATISSE data and reordered the GRAVITY closure phase telescope triplets to match the order of the captions. The changes in closure phase sign from reordering were taken into account.}
    \label{fig:polycf}
\end{figure*}

\begin{figure*}
    \centering
    \includegraphics[width=\textwidth]{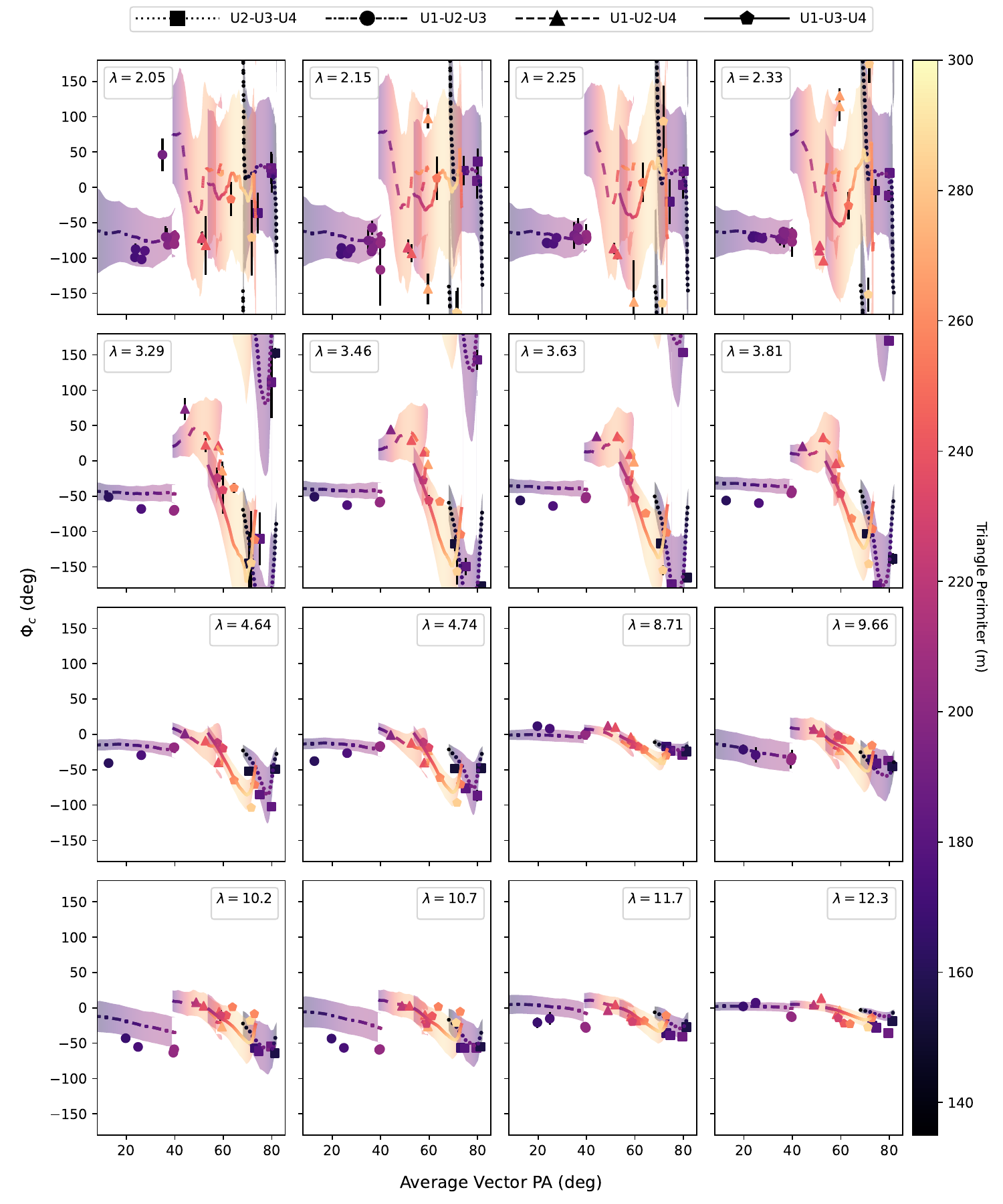}
    \caption{Closure phase of the median polychromatic model per wavelength with 1$\,\sigma$ errors compared to observations. The closure phases were calculated for each of the four telescope triplets at all accessible \textit{uv} positions for the UTs where NGC\,1068 has a minimum altitude of 35$^\circ$. The average vector PA is the average of the PA for each of the three baselines in the triplet ensuring the telescopes are in ascending numerical order. E.g. the average PA for U1-U2-U3 is the is average of the PA of the baselines U1-U2, U2-U3, and U1-U3 {not U3-U1}.  For visual purposes, we have averaged the different BCD positions for the MATISSE data and reordered the GRAVITY closure phase telescope triplets to match the order of the captions. The changes in closure phase sign from reordering were taken into account.} 
    \label{fig:polycfmed}
\end{figure*}

\begin{figure*}
    \centering
    \includegraphics[width=0.7\textwidth]{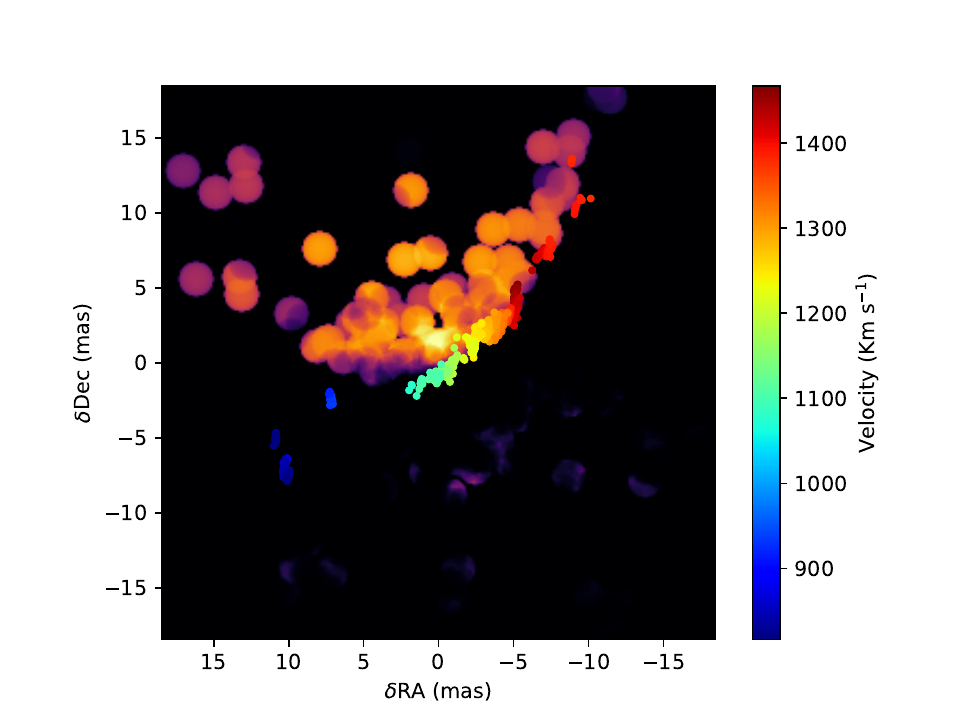}
    \caption{Comparison of the H$_2$O maser emission from \citet{gallimore_high-sensitivity_2023} to the unconvolved polychromatic model at 2$\,\mu$m. The model has a log colour scaling to highlight faint structure. The model centre is aligned with the 5\,GHz position of S1 from their work. The velocity is their recession velocity.}
    \label{fig:masers}
\end{figure*}

\begin{figure*}
    \centering
    \includegraphics[width=0.9\textwidth]{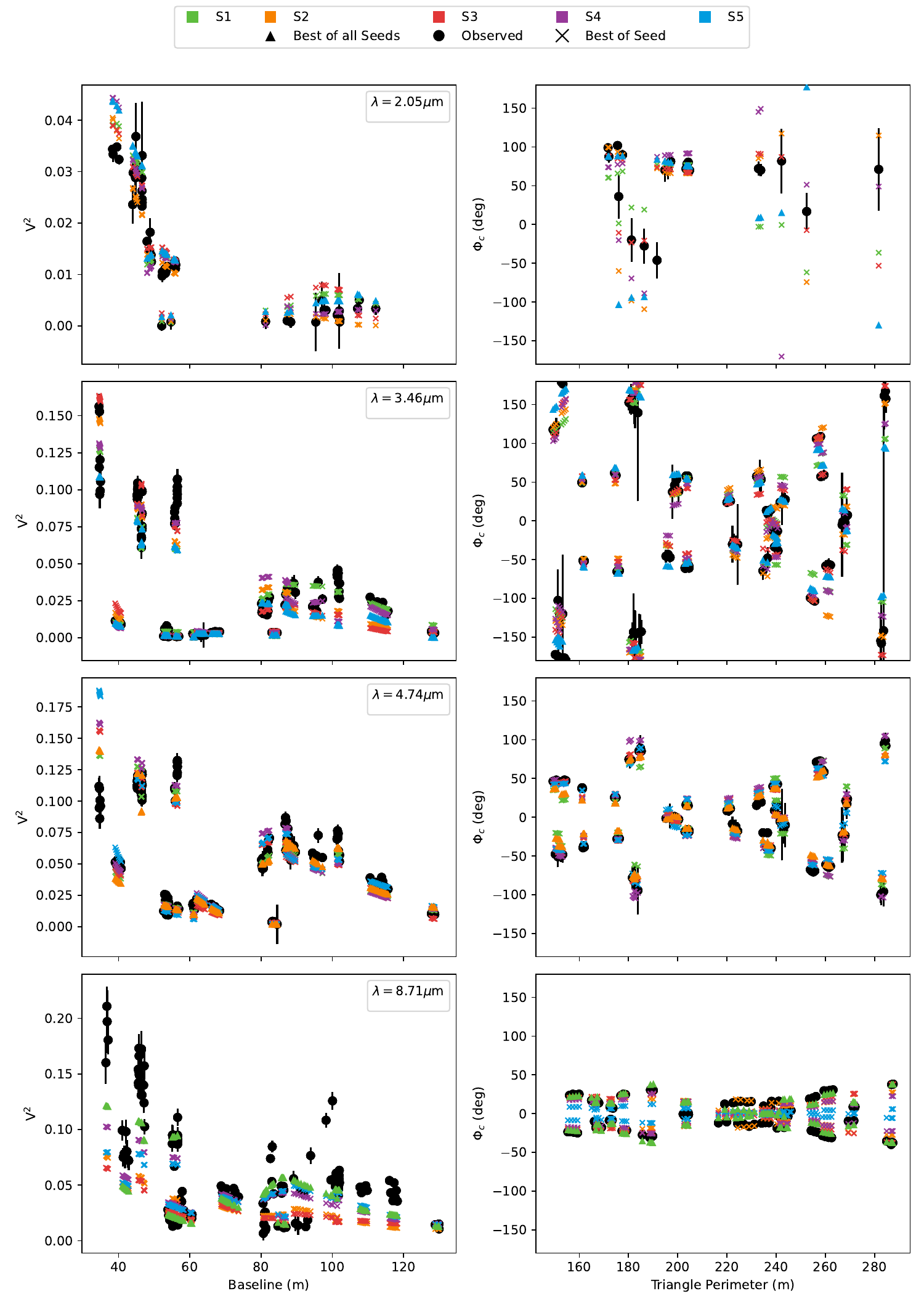}
    \caption{V$^2$ and $\Phi_c$ from the best models from Sect.\,\ref{sec:endind} at one wavelength per band for each seed compared to the observed data.}
    \label{fig:observables_initial}
\end{figure*}

\begin{figure*}
    \centering
    \includegraphics[width=0.9\textwidth]{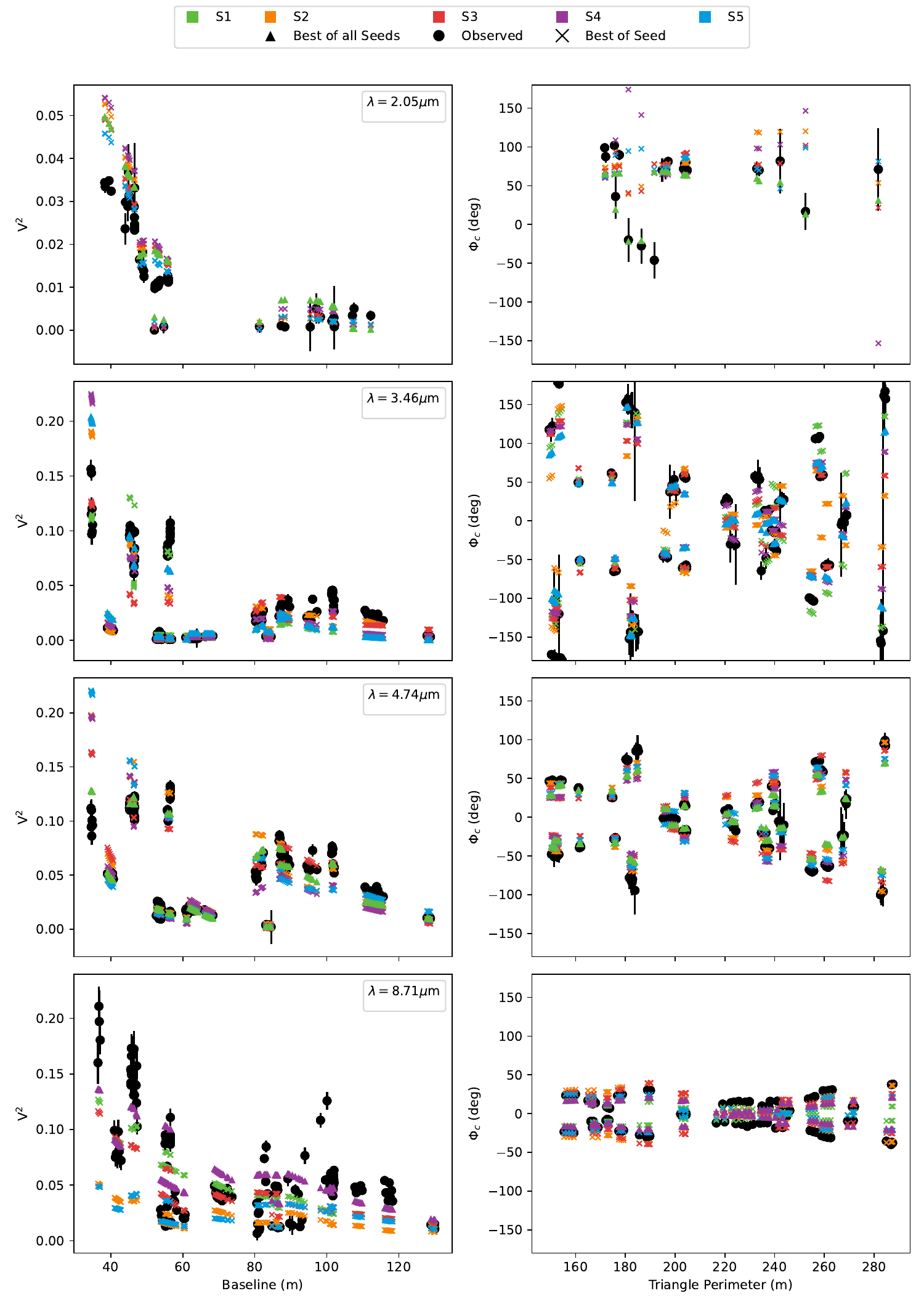}
    \caption{V$^2$ and $\Phi_c$ from the best models in Sect.\,\ref{sec:faceind} at one wavelength per band for each seed compared to the observed data.}
    \label{fig:observables_initial_face}
\end{figure*}

\begin{figure*}
    \centering
    \includegraphics[width=\textwidth]{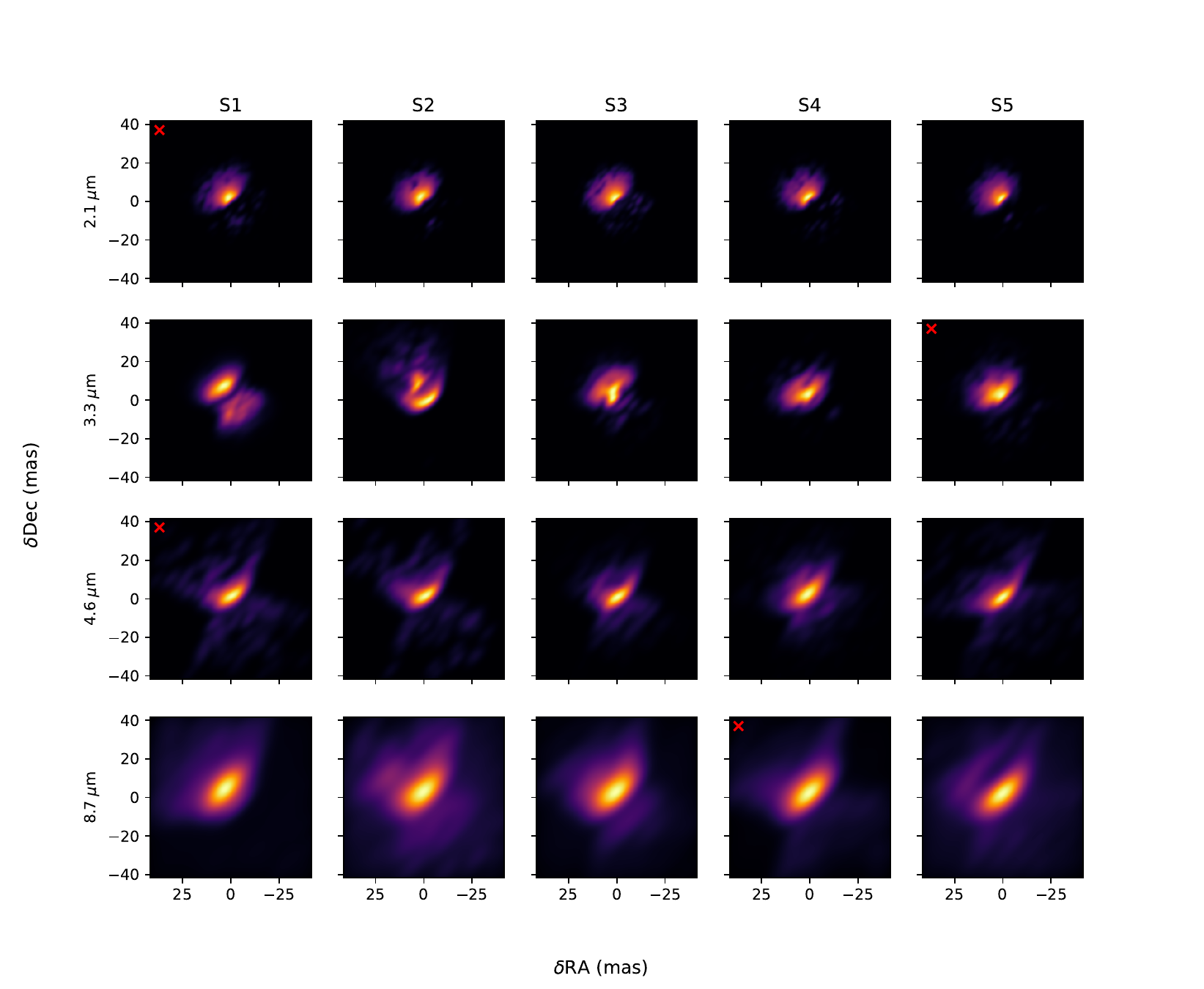}
    \caption{Best model for each band (row) and each seed (column) for the initial Sy1 band-by-band fit as determined by the maximum likelihood. The best seed for each band is marked with a red x at the centre. The K-,L-,M-,N-bands are evaluated at 2.1\,$\mu$m,3.5\,$\mu$m,4.7\,$\mu$m, and 8.5\,$\mu$m, respectively. All images are normalised and given a 0.6 power colour scaling to match that of \citet{gamez_rosas_thermal_2022} for easy comparison.}
    \label{fig:bbb_initial_face}
\end{figure*}

\begin{figure*}
    \centering
    \includegraphics[width=\textwidth]{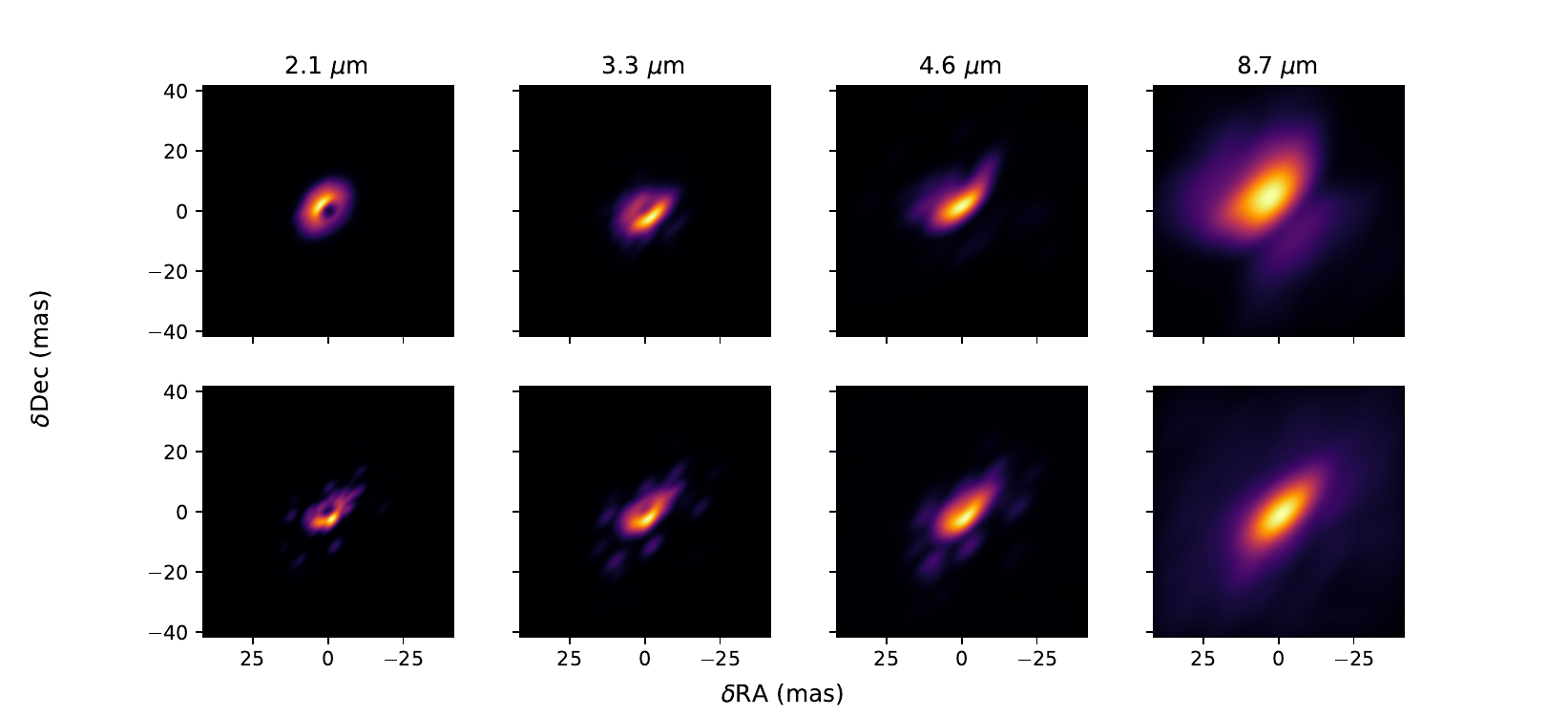}
    \caption{Best model for the initial Sy1 ring-case band-by-band (top row) and polychromatic (bottom row) fit as determined by the maximum likelihood. All images are normalised and given a 0.6 power colour scaling to match that of \citet{gamez_rosas_thermal_2022} for easy comparison.}
    \label{fig:ringmodel}
\end{figure*}

\begin{figure*}
    \centering
    \includegraphics[width=0.85\textwidth]{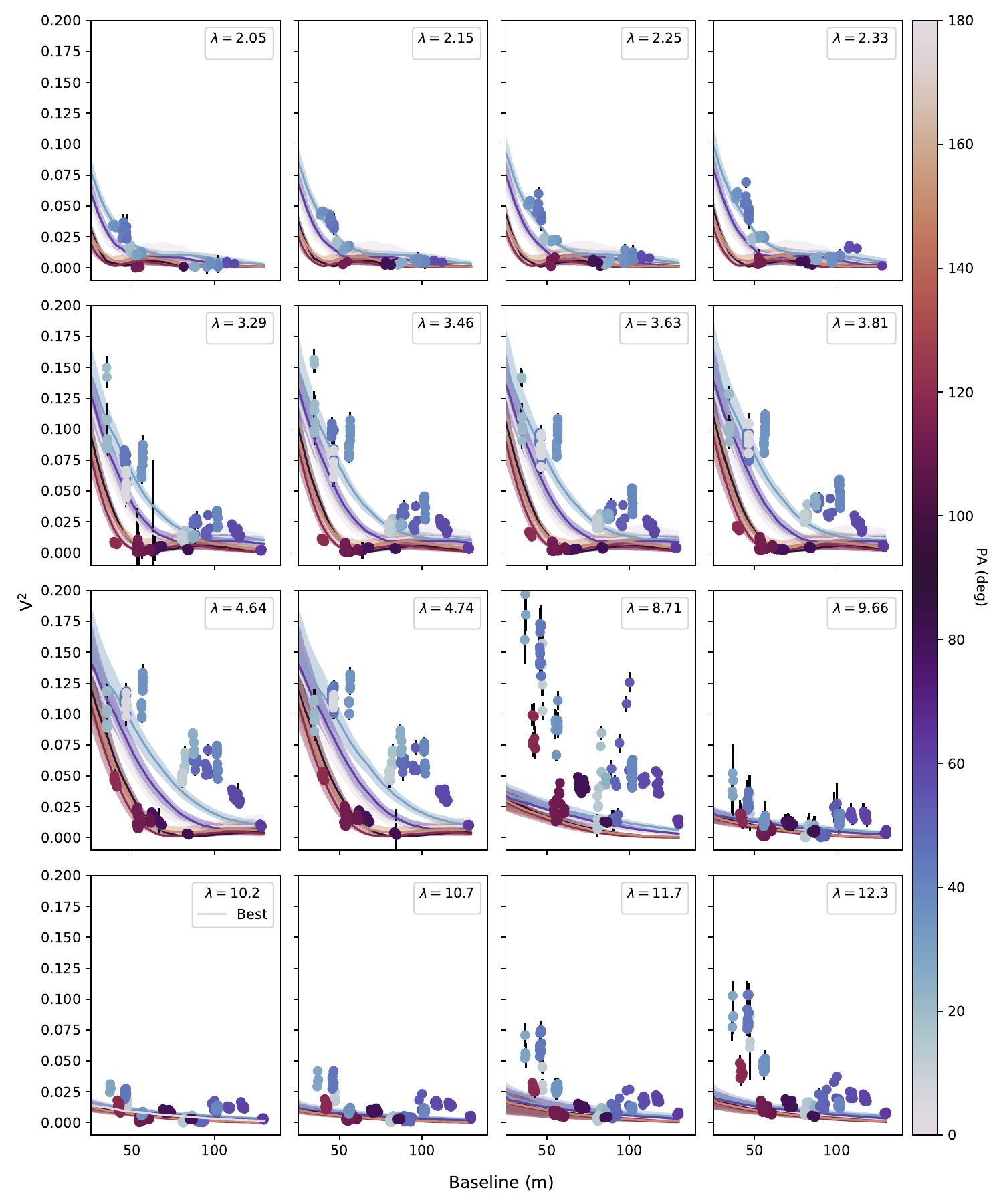}
    \caption{Same as Fig.\ref{fig:polyv2med} for the ring case of Sect.\ref{sec:ring}. The line is the median fit of the polychromatic model with 1$\,\sigma$ errors as the shaded region evaluated at different PA.}
    \label{fig:ringmodelv2}
\end{figure*}

\begin{figure*}
    \centering
    \includegraphics[width=\textwidth]{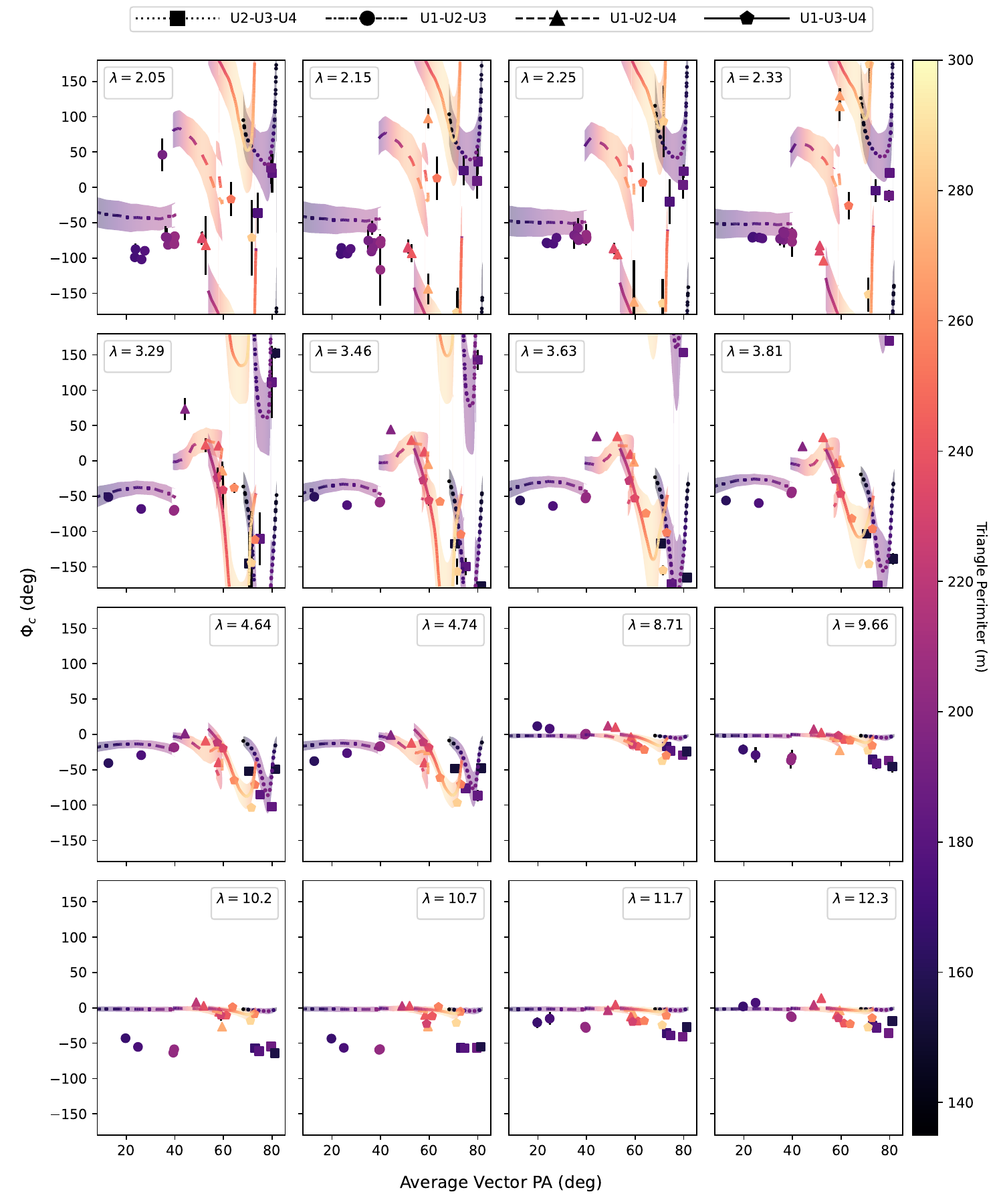}
    \caption{Same as Fig.\ref{fig:polycfmed} for the ring case of Sect.\ref{sec:ring}. The closure phases were calculated for each of the four telescope triplets at all accessible \textit{uv} positions for the UTs where NGC\,1068 has a minimum altitude of 35$^\circ$. The average vector PA is the average of the PA for each of the three baselines in the triplet ensuring the telescopes are in ascending numerical order. E.g. the average PA for U1-U2-U3 is the is average of the PA of the baselines U1-U2, U2-U3, and U1-U3 {not U3-U1}.  For visual purposes, we have averaged the different BCD positions for the MATISSE data and reordered the GRAVITY closure phase telescope triplets to match the order of the captions. The changes in closure phase sign from reordering were taken into account.} 
    \label{fig:ringmodelphi}
\end{figure*}

\end{appendix}

\end{document}